\documentclass[lettersize,journal]{IEEEtran}

\usepackage{makecell}
\usepackage{caption} 
\usepackage{graphicx} 
\usepackage{subfigure}
\usepackage{amsmath,amsfonts}
\usepackage{algorithmic}
\usepackage{algorithm}
\usepackage{array}
\usepackage[caption=false,font=normalsize,labelfont=sf,textfont=sf]{subfig}
\usepackage{textcomp}
\usepackage{stfloats}
\usepackage{url}
\usepackage{verbatim}
\usepackage{graphicx}
\usepackage{cite}
\usepackage{amssymb}
\usepackage{booktabs}
\usepackage{threeparttable}
 \usepackage{tabularx}
\usepackage{multirow}
\usepackage{multicol}

\newtheorem{theorem}{Theorem}
\newtheorem{definition}{Definition}
\begin{document}

\title{Quaternion Tensor  Completion with Sparseness for Color Video Recovery }
\author{Liqiao Yang, Kit Ian Kou, Jifei Miao, Yang Liu
	\IEEEmembership{Senior Member, IEEE}, Maggie Pui Man Hoi
	\thanks{Liqiao Yang is with the Department of  Mathematics, Faculty of Science and Technology, University of Macau, Macau, 999078, China (e-mail: liqiaoyoung@163.com)}
	\thanks{Kit Ian Kou is with the Department of  Mathematics, Faculty of  Science and Technology, University of Macau, Macau, 999078, China (e-mail: kikou@umac.mo)}
	\thanks{Jifei Miao is with the School of Mathematics and Statistics, Yunnan University, Kunming, Yunnan, 650091, China (e-mail: jifmiao@163.com)}
	\thanks{Yang Liu is with the College of Mathematics and Computer Science and the College of Mathematical Medicine, Zhejiang Normal University, Jinhua 321004, China (e-mail: liuyang@zjnu.edu.cn)}
	\thanks{Maggie Pui Man Ho is with the State Key Laboratory of Quality Research in Chinese Medicine, Institute of Chinese Medical Sciences, University of Macau, Macau, 999078, China (e-mail:  MagHoi@um.edu.mo)}}

\markboth{Journal of \LaTeX\ Class Files,~Vol.~14, No.~8, August~2021}%
{Shell \MakeLowercase{\textit{et al.}}: A Sample Article Using IEEEtran.cls for IEEE Journals}


\maketitle

\begin{abstract}
A novel low-rank completion algorithm based on the quaternion tensor is proposed in this paper. This approach uses the TQt-rank of quaternion tensor to maintain the structure of  RGB channels throughout the entire process. In more detail, the pixels in each frame are encoded on three imaginary parts of a quaternion as an element in a quaternion matrix. Each quaternion matrix is then stacked into a quaternion tensor. A logarithmic function and truncated nuclear norm are employed to characterize the rank of the quaternion tensor in order to promote the low rankness of the tensor. Moreover, by introducing a newly defined quaternion tensor discrete cosine transform-based (QTDCT) regularization to the low-rank approximation framework, the optimized recovery results can be obtained in the local details of color videos. In particular, the sparsity of the quaternion tensor is reasonably characterized by $l_1$ norm in the QDCT domain. This strategy is optimized via the two-step  alternating direction method of multipliers (ADMM) framework. Numerical experimental results for recovering color videos show the obvious advantage of the proposed method over other potential competing approaches.
\end{abstract}


\section{Introduction}
\IEEEPARstart{W}{ITH} the development of  computer vision processing, the effective recovery of missing color videos has become an indispensable research direction because high-dimensional images  would inevitably be damaged; for instance, because of a bad imaging environment or poor transfer process. The main task of color video restoration is to recover missing elements in incomplete color video from limited observation. As a kind of  high-dimensional data, color videos can be naturally represented by tensors because tensors are high-dimensional vectors, which are the generalization of matrices and vectors. Based on tensor representation,  color video restoration can be transformed into a tensor completion (TC) problem, and the key to this issue is to ascertain the correlation between observed and missing elements \cite{DBLP:journals/pami/LiuMWY13}.

Like matrix completion, comprehensive research has been conducted on low-rank based models, resulting in numerous improvements. Concretely, high-dimensional visual data has many similar and repeating patterns, thus, the image or video data has a highly correlated underlying structure \cite{DBLP:conf/cvpr/ZhangEAHK14}. In this way, the main task of low-rank TC (LRTC) is to build a low-rank regularization as per the previous description of  the association between the observed and missing elements, enabling the filling of missing elements. 

Unlike matrix-based methods, which need to pull high-dimensional data into vectors or matrices for processing, tensor can maintain the high-dimensional structure of the data itself. For example, a color image (with red, green, and blue channels) can be regarded as a third-order tensor, and a color video consisting of frame-by-frame color images can be considered as a fourth-order tensor similarly. Furthermore, unlike the uniqueness of the definition of the rank of a matrix, the definition of the rank of a tensor is not unique,\textit{ e.g.}, the CANDECOMP/PARAFAC (CP) rank \cite{DBLP:journals/tsp/YokotaZC16 , DBLP:journals/pami/ZhaoZC15} 
, the Tucker rank \cite{gandy2011tensor} and the tubal rank that in the light of tensor singular value decomposition (t-SVD) \cite{DBLP:journals/tsp/ZhangA17, DBLP:journals/siammax/KilmerBHH13, DBLP:conf/icpr/XueQLJ18}, \textit{etc}.

The CP rank and the Tucker rank are the two most commonly utilized tensor ranks. The CP rank is defined as the minimum number of rank-1 tensors used to express a tensor by tallying this series of rank-1 tensors. It can be observed that although the CP rank is consistent with the definition of rank of matrix, it is difficult to identify a replaceable form that can be solved \cite{DBLP:journals/jacm/HillarL13}. Hence, many CP-based methods directly utilize CP decomposition to characterize low-rankness of high-dimensional visual data \cite{DBLP:journals/tsp/YokotaZC16}. The Tucker rank is defined as a vector, the k-th element in the vector is the rank of the mode-k unfolding of the tensor that is an unfolding matrix \cite{kolda2009tensor}.  Inspired by the success of nuclear norm (NN) as the tightest convex surrogate of the matrix rank, methods based on Tucker rank generally adopt the sum of NN (SNN) of all the unfolding matrices of the tensor. Consequently, the problem evolves to how to optimize the SNN to achieve low-rank characterization. To avoid all singular values being treated equally, the method of adding weights to the NN of unfolding matrices was proposed in \cite{DBLP:journals/pami/LiuMWY13}. The weighted method entails introducing additional parameters, which weakens the degree of robustness. To overcome this disadvantage, the logarithm of singular values was designed, which allows larger singular values to be given smaller weights, because larger singular values carry more important information 
\cite{ji2017non}. Although SNN can effectively apply the relationship among different modes after unfolding, the structure of high-dimensional data is inevitably destroyed during the process of unfolding a tensor to a matrix along one mode. 

\begin{figure} [htbp]
	\centering
	\includegraphics[width=90mm]{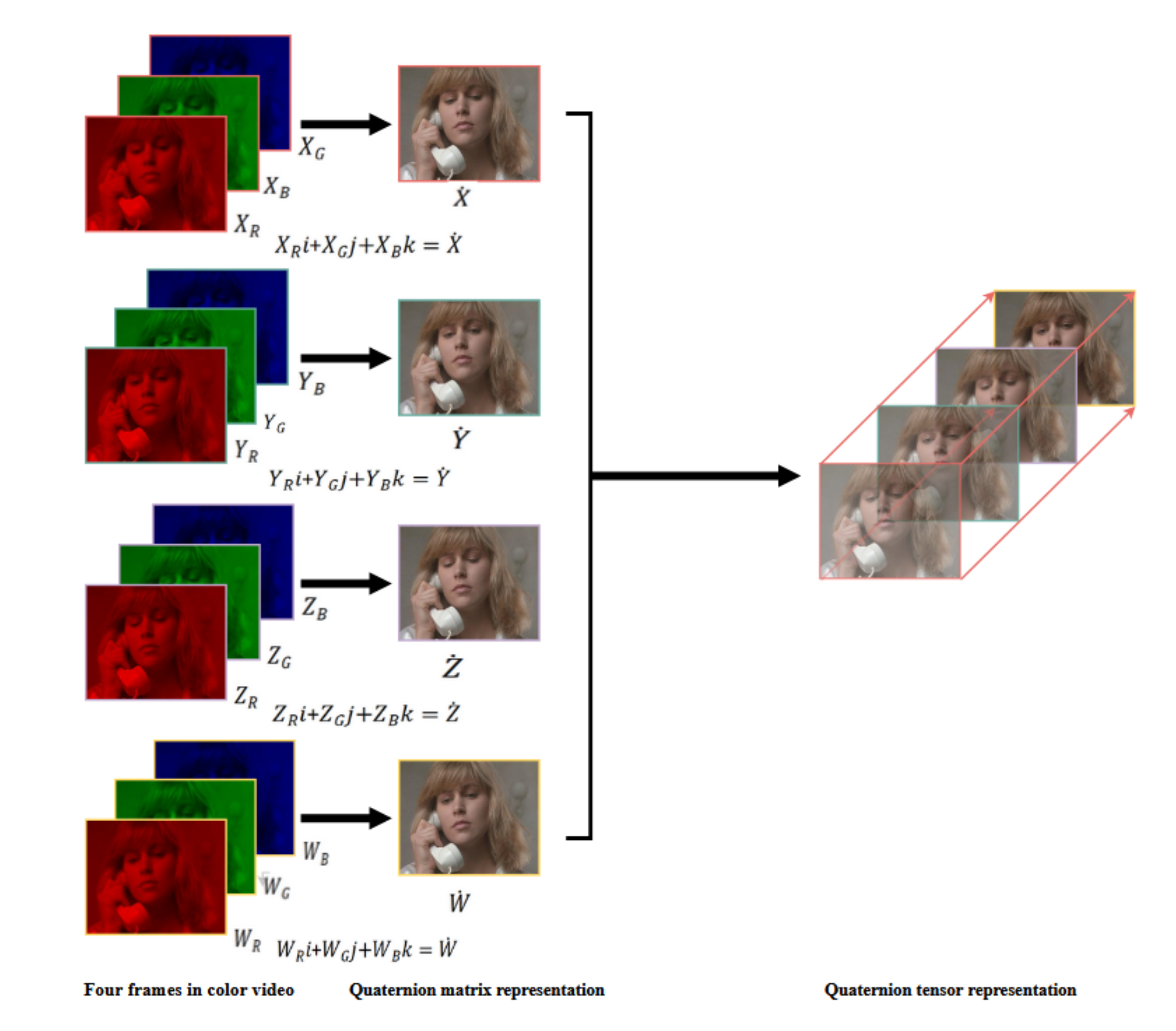}
	\caption{The illustration of a color video is represented by a three-dimensional quaternion tensor, taking four frames as an example.}
	\label{f1}
\end{figure}

Moreover, based on t-SVD, the tubal rank also has been extensively studied in recent years. The definition of t-SVD was suggested in \cite{DBLP:journals/siammax/KilmerBHH13}, then based on the novel definition of tensor-tensor product and some properties of t-SVD that are similar to SVD, the definition of the tubal nuclear norm was put forward \cite{DBLP:journals/tsp/ZhangA17}. This norm was defined by the sum of NN of  all slices of tensor in the Fourier domain. Furthermore, in  \cite{DBLP:conf/icpr/XueQLJ18}, the tensor truncated nuclear norm (T-TNN) was proposed for a superior estimation of rank since the larger the singular value, the less it will affect the rank. In addition, in \cite{DBLP:conf/cvpr/Lu0W19}, inspired by a novel tensor-tensor product defined in \cite{kernfeld2015tensor}, a more general tensor NN was proposed induced by linear transforms such as Discrete Cosine Transform (DCT) and Random Orthogonal Matrix (ROM), \textit{etc}. Approaches based on the tubal NN  have proven to be effective in practical application by preserving the intrinsic structure of a tensor \cite{DBLP:journals/tnn/HuTZXY17, DBLP:journals/tip/ZhouLLZ18}. However, the definition of  t-SVD is designed for third-order tensors, so when dealing with color videos, this type of method needs to perform dimensional reduction operations, which also means that the structure among RGB will be disregarded. To overcome the above issue, in this paper, a color video is represented by a third order quaternion tensor as the pixel values of RGB channels can be placed in three imaginary parts of a quaternion. In fact, the quaternion-based strategies were proposed to process high-dimensional visual data and outstanding results were returned, e.g., color face recognition \cite{DBLP:journals/tip/ZouKW16}, color image and video completion \cite{DBLP:journals/pr/MiaoKL20}, classification of biomedical color image \cite{DBLP:journals/mta/AmakdoufZMTCQ21}, \textit{etc}.  In most of the quaternion-based methods, a color image is regarded as a quaternion matrix. In order to extend the definition of  quaternion matrix and fully preserve the structure of  color channels, in  \cite{DBLP:journals/pr/MiaoKL20}, the concept of quaternion tensor is presented. More specifically, a color image and a color video can be  reconstructed as a two-dimensional quaternion tensor and a three- dimensional quaternion tensor, respectively. Considering the definition of quaternion tensor, in this paper, a color video is put in a three- dimensional quaternion tensor, and the visible explanation of a three- dimensional quaternion tensor representing a color video can be seen in Fig.\ref{f1}. 

As shown in Fig.\ref{f1}, it is expected to extend the t-SVD to the quaternion domain such that the color information and spatial structure of a video are preserved to the greatest extent. Not only that, inspired by the novel transform-based tensor product and corresponding transform-based quaternion tensor SVD (TQt-SVD) defined in the quaternion domain in \cite{DBLP:journals/corr/abs-2112-08921}, a TQt-SVD based on Quaternion Discrete Cosine Transform (QDCT) is adopted to address the low-rank recovery problem in this paper. Furthermore, to more accurately depict the low-rankness, the proposed model encompasses a truncated strategy and logarithmic function of singular values.

Not limited to the low-rank property, due to the fact that some signals in real visual data, such as audio, color images, and videos, can be represented sparsely by fixed bases \cite{wright2010sparse}, the sparse property of a color video can be found in some transformed domain. Thus, many methods previously considered sparseness by formulating it as a $l_1$ norm regularizer  in a specific domain \cite{DBLP:journals/spic/DongXGHW18, DBLP:journals/nla/WangLC21}. 

Considering the above, this paper develops a novel quaternion tensor recovery model, which accounts for both the low-rankness and sparseness of the underlying quaternion tensor. The detailed demonstration of this regularization can be found in Section \ref{Main1}.

The main contribution of this paper is summarized as follows:

 \begin{itemize}
 	\item{This work concentrate on quaternion tensor recovery, and a novel quaternion tensor completion model is developed, which simultaneously depicts the low rankness and the sparseness of the target quaternion tensor. The color visual data is addressed in the quaternion domain, which prevents corruption of the data structures.}
 	\item{The low-rankness is depicted by introducing a truncated strategy and logarithmic function to the singular value for a more accurate substitute of the rank function, which is based on TQt-SVD.}
 	\item{There is no decrease in dimensionality  or slicing of the quaternion tensor, and the sparse regularization of the target quaternion tensor is designed, which is based on the  quaternion tensor DCT (QTDCT) defined in this paper.}
 	\item{This model alternates between the two steps with ADMM so as to optimize its development. The simulations on color videos recovery demonstrate that the proposed method outperforms state-of-the-art and comparable tensor-based and quaternion tensor-based methods.}
 \end{itemize} 

The rest of this paper is arranged as follows. Section \ref{P} presents some notations preliminaries used in this paper. Section \ref{Main1} is devoted to introduce the proposed low-rank approximation and the sparse regularization. Section \ref{Main2} gives the corresponding recovery models and two-step optimization with ADMM. Section \ref{E} then provides the numerical results to show the efficiency of the proposed method, while Section \ref{C} offers a conclusion to the work.

\section{Notations and Preliminaries}
\label{P}
\subsection{Notations}
In the real domain $\mathbb{R}$, scalar, vector,  matrix, and tensor are denoted as a, $\mathbf{a}$, $\mathbf{A}$, and $\mathcal{A}$, respectively. In the quaternion domain $\mathbb{H}$, scalar, vector, matrix, and tensor are denoted as $\dot{a}$, $\dot{\mathbf{a}}$, $\dot{\mathbf{A}}$, and $\dot{\mathcal{A}}$, respectively. Besides, the complex space is denoted as $\mathbb{C}$. For a quaternion $\dot{q}$,  the real part and imaginary part are denoted as  $\mathfrak{R}(\dot{q})$ and $\mathfrak{I}(\dot{q})$. For a quaternion tensor $\dot{\mathcal{A}}\in\mathbb{H}^{I_1\times I_2\times \cdots \times I_N}$, the frontal slice is denoted by the Matlab notation $\dot{\mathcal{A}}(:, :, i_3, \cdots, i_N)$, where $i_k=1, \cdots, I_k,$ for $k=3, \cdots, N$. Compactly, $\dot{\mathcal{A}}(:, :, i_3, \cdots, i_N)$ is denoted by $\dot{\mathcal{A}}^{(k)}$. The transpose, conjugate transpose, and inverse are denoted as $\mathbf{(\cdot)}^T$, $\mathbf{(\cdot)}^H$, and $\mathbf{(\cdot)}^{-1}$, respectively. The inner product of $*_1$ and $*_2$ is defined as $\langle*_1\cdot*_2\rangle \triangleq \text{tr}(*_1^H*_2)$, where $tr(\cdot)$ is the trace function. The $\times_k$ represents the $k$-mode product. The trace function of $\dot{\mathcal{A}}$ is defined as $tr(\dot{\mathcal{A}})=\sum_{i=1}^{I_3}tr(\dot{\mathcal{A}}^{(i)}).$ $\mathbf{I}$ denotes the identity matrix. 
\subsection{Preliminaries}
\subsubsection{Quaternion basics}

As an extension of complex numbers, quaternions are discovered by Hamilton in 1843 \cite{doi:10.1080/14786444408644923}. A quaternion number $\dot{q}\in\mathbb{H}$  can be represented as 
\begin{equation*}
\dot{q}=q_0+q_1\emph{i}+q_2\emph{j}+q_3\emph{k},
\end{equation*}
where $q_n\in\mathbb{R}$ $(n=0,1,2,3)$, and \emph{i, j, k} are imaginary units which have the following relations:
\begin{equation*}
\begin{cases}
\emph{i}^2= \emph{j}^2 =\emph{k}^2= \emph{i}\emph{j}\emph{k}=-1\\
\emph{i}\emph{j}=-\emph{j}\emph{i} = \emph{k},    \emph{j}\emph{k}=-\emph{k}\emph{j} = \emph{i},  \emph{k}\emph{i}=-\emph{i}\emph{k} = \emph{j}.
\end{cases}
\end{equation*}
Let $\dot{q}=\mathfrak{R}(\dot{q})+\mathfrak{I}(\dot{q})$, where $\mathfrak{R}(\dot{q}) \triangleq q_0$ is the real part of  $\dot{q}$, and $\mathfrak{I}(\dot{q}) \triangleq q_1\emph{i}+q_2\emph{j}+q_3\emph{k}$ is the  imaginary part of $\dot{q}$. If real part $q_0 = 0$, $\dot{q}$ is a pure quaternion. A quaternion can be denoted by the Cayley-Dickson notation: $\dot{q}=\alpha+\beta j$, where $\alpha=a+bi$ and $\beta=c+di \in\mathbb{C}$. The conjugate and the modulus of $\dot{q}$  are formulated as $\dot{q}^{*} \triangleq q_0-q_1\emph{i}-q_2\emph{j}-q_3\emph{k}$  and $|\dot{q}|\triangleq \sqrt{\dot{q}\dot{q}^{*}}=\sqrt{q_0^2+q_1^2+q_2^2+q_3^2}$, respectively. Given two quaternions $\dot{p}$ and $\dot{q}\in\mathbb{H}$,  the addition and multiplication are  separately formulated as 
\begin{equation}\nonumber 
\dot{p}+\dot{q}=(p_0+q_0)+(p_1+q_1)\emph{i}+(p_2+q_2)\emph{j}+(p_3+q_3)\emph{k},
\end{equation}
\begin{equation}\nonumber
\begin{aligned}
\dot{p}\dot{q}=& (p_0q_0-p_1q_1-p_2q_2-p_3q_3)\\&+(p_0q_1+p_1q_0+p_2q_3-p_3q_2)\emph{i}
\\&+(p_0q_2-p_1q_3+p_2q_0+p_3q_1)\emph{j}\\&+(p_0q_3+p_1q_2-p_2q_1+p_3q_0)\emph{k}.
\end{aligned}
\end{equation}
Note that the multiplication in the quaternion domain is not commutative, \textit{i.e.}, $\dot{p}\dot{q} \neq \dot{p}\dot{q}$.

A quaternion matrix $\dot{\mathbf{Q}}=(\dot{q}_{ij})\in\mathbb{H}^{M \times N}$ is formulated as $\dot{\mathbf{Q}}=\mathbf{Q}_0+\mathbf{Q}_1\emph{i}+\mathbf{Q}_2\emph{j}+\mathbf{Q}_3\emph{k}$, where $\mathbf{Q}_n\in\mathbb{R}^{M \times N}$ $(n=0,1,2,3)$ are real matrices. When  $\mathbf{Q}_0 = \mathbf{0}$, $\dot{\mathbf{Q}}$ is pure quaternion matrix. 

\begin{definition}[\textbf{The Cayley-Dickson form}\label{def1} \cite{DBLP:journals/sigpro/BihanM04}]
 Let $\dot{\mathbf{Q}}=\mathbf{Q}_0+\mathbf{Q}_1\emph{i}+\mathbf{Q}_2\emph{j}+\mathbf{Q}_3\emph{k}\in\mathbb{H}^{M \times N}$, the Cayley-Dickson form of $\dot{\mathbf{Q}}$ is $\dot{\mathbf{Q}}=\mathbf{Q}_p+\mathbf{Q}_q\emph{j}$, where $\mathbf{Q}_p=\mathbf{Q}_0+\mathbf{Q}_1\emph{i}$ and $\mathbf{Q}_q=\mathbf{Q}_2+\mathbf{Q}_3\emph{i}\in\mathbb{C}^{M \times N}$. Besides, the isomorphic complex  matrix representation of  $\dot{\mathbf{Q}}$ is given by  $\mathbf{Q}_c\in\mathbb{C}^{2M \times 2N}$:
\begin{equation*}\label{ct}
\mathbf{Q}_c={
	\left( \begin{array}{cc}
	\mathbf{Q}_p & \mathbf{Q}_q  \\
	-\mathbf{Q}_q^* & \mathbf{Q}_p^*\\
	\end{array}
	\right )_{2M \times 2N}}.
\end{equation*}
\end{definition}

\begin{definition}[\textbf{The rank of quaternion matrix} \cite{zhang1997quaternions}]Let  $\dot{\mathbf{Q}}=(\dot{q}_{ij})\in\mathbb{H}^{M \times N}$, the rank of $\dot{\mathbf{Q}}$ is defined as  the  maximum number of right (left) linearly independent columns (rows)  of  $\dot{\mathbf{Q}}$.
\end{definition}

\begin{theorem}[\textbf{Quaternion SVD (QSVD)} \cite{zhang1997quaternions}]
	\label{th1}
	Given $\dot{\mathbf{Q}}\in\mathbb{H}^{M \times N}$ be of rank $r$. There are two unitary quaternion matrices $\dot{\mathbf{U}}\in\mathbb{H}^{M \times M}$
	and $\dot{\mathbf{V}}\in\mathbb{H}^{N \times N}$ such that
	\begin{equation*}
	\dot{\mathbf{Q}}=\dot{\mathbf{U}}
	\left( \begin{array}{cc}
	\mathbf{\Sigma}_r & \mathbf{0}  \\
	\mathbf{0} & \mathbf{0}\\
	\end{array}
	\right )\dot{\mathbf{V}}^H= \dot{\mathbf{U}}\mathbf{\Lambda}\dot{\mathbf{V}}^H,
	\end{equation*}
	where $\mathbf{\Sigma}_r=diag({\sigma_1,\cdots, \sigma_r})\in\mathbb{R}^{r\times r}$is a real diagonal matrix, and singular values $\sigma_i$ are positive, where $i=1,\cdots,r$. 
\end{theorem}

Similar to the case of the real matrix, the Frobenius norm and $l_1$ norm of  the quaternion matrix $\dot{\mathbf{Q}}=(\dot{q}_{ij})\in\in\mathbb{H}^{M \times N}$ are separately formulated as  $\parallel\dot{\textbf{Q}}\parallel_F = \sqrt{\sum_{i=1}^{M}\sum_{j=1}^{N}|\dot{q}_{ij}|^2}=\sqrt{tr(\dot{\textbf{Q}}^H\dot{\textbf{Q}})}$, $\parallel\dot{\textbf{Q}}\parallel_1 = \sum_{i=1}^{M}\sum_{j=1}^{N}|\dot{q}_{ij}|$. The rank of $\dot{\mathbf{Q}}$ is the $l_0$ norm of the vector $\{\sigma_i(	\dot{\mathbf{Q}})\}_{i=1}^{\min(M, N)}$. Since the $l_0$ norm is nonconvex, quaternion NN is developed as $\parallel\dot{\mathbf{Q}}\parallel_*=\sum_{i=1}^{\min(M,N)}\sigma_{i}(\dot{\mathbf{Q}})$. 
\subsubsection{Quaternion tensor basics}
\begin{definition}[\textbf{Quaternion tensor}\cite{DBLP:journals/pr/MiaoKL20}]
	\label{def3}
	A multi-dimensional array or an Nth-order tensor is defined as a quaternion tensor when its entries are quaternions. Specifically, quaternion tensor $\dot{\mathcal{T}}=(\dot{t}_{i_1i_2\ldots i_N})\in\mathbb{H}^{I_1\times I_2\times \cdots \times I_N}$ is formulated as $\dot{\mathcal{T}}=\mathcal{T}_0+\mathcal{T}_1\emph{i}+\mathcal{T}_2\emph{j}+\mathcal{T}_3\emph{k}$, where $\mathcal{T}_n\in\mathbb{R}^{I_1\times I_2\times \cdots \times I_N}$ $(n=0,1,2,3)$ are real tensors. When  $\mathcal{T}_0 $ is a zero tensor, $\dot{\mathcal{T}}$ is pure quaternion tensor.  
\end{definition}

As introduced above, in \cite{DBLP:journals/corr/abs-2112-08921}, a general quaternion tensor product and the corresponding TQt-SVD are proposed. 

\begin{definition}[\textbf{$\star_{QT}$-product} \cite{DBLP:journals/corr/abs-2112-08921}]
	\label{def4}
Given two Nth-order ($N\geq3$) quaternion tensors $\dot{\mathcal{A}}\in\mathbb{H}^{I_1\times l\times I_3\times \ldots \times I_N}$, $\dot{\mathcal{B}}\in\mathbb{H}^{l\times I_2\times I_3\times \ldots \times I_N}$ and $N-2$ invertible quaternion matrices $\dot{\mathbf{Q}}_3\in\mathbb{H}^{I_3\times I_3}, \cdots,  \dot{\mathbf{Q}}_N\in\mathbb{H}^{I_N\times I_N}$, the $\star_{QT}$-product is defined as 
\begin{equation}
	\dot{\mathcal{T}}=\dot{\mathcal{A}}\star_{QT}\dot{\mathcal{B}}=(\hat{\dot{\mathcal{A}}}\star_{QF}\hat{\dot{\mathcal{B}}})\times_3\dot{\mathbf{Q}}_3^{-1}\times_4 \cdots \times_N\dot{\mathbf{Q}}_N ^{-1},
\end{equation}	
where $\hat{\dot{\mathcal{A}}}=\dot{\mathcal{A}}\times_3\dot{\mathbf{Q}}_3\times_4 \cdots \times_N\dot{\mathbf{Q}}_N $ and $\hat{\dot{\mathcal{B}}}=\dot{\mathcal{B}}\times_3\dot{\mathbf{Q}}_3\times_4 \cdots \times_N\dot{\mathbf{Q}}_N $. The $\star_{QF}$-product is the quaternion facewise produce, \textit{i.e.}, $\dot{\mathcal{F}}=\dot{\mathcal{A}}\star_{QF}\dot{\mathcal{B}}$ such that the frontal slice of $\dot{\mathcal{F}}$ satisfies $\dot{\mathcal{F}}(:, :, i_3, \cdots, i_N)=\dot{\mathcal{A}}(:, :, i_3, \cdots, i_N)\dot{\mathcal{B}}(:, :, i_3, \cdots, i_N)$.
\end{definition}

Based on $\star_{QT}$-product, the TQt-SVD is proposed.
\begin{theorem}[\textbf{TQt-SVD }\cite{DBLP:journals/corr/abs-2112-08921}]
	\label{th2}
Given $\dot{\mathcal{T}}\in\mathbb{H}^{I_1\times I_2\times \cdots \times I_N}$. There are two unitary quaternion tensors $\dot{\mathcal{U}}\in\mathbb{H}^{I_1\times I_1\times \cdots \times I_N}$.
and $\dot{\mathcal{V}}\in\mathbb{H}^{I_2\times I_2\times \cdots \times I_N}$. such that
\begin{equation*}
	\dot{\mathcal{T}}= \dot{\mathcal{U}}\star_{QT}\dot{\mathcal{D}}\star_{QT}\dot{\mathcal{V}}^H,
\end{equation*}
where $\dot{\mathcal{D}}\in\mathbb{H}^{I_1\times I_2\times \cdots \times I_N}$ is an f-diagonal quaternion tensor which means that only its frontal slices have non-zero elements on them. 
\end{theorem}

Further,  the definition of Nth-order quaternion tensor rank (TQt-rank) was proposed.

\begin{definition}[\textbf{TQt-rank} \cite{DBLP:journals/corr/abs-2112-08921}] Let  $\dot{\mathcal{T}}\in\mathbb{H}^{I_1\times I_2\times \cdots \times I_N} (N\geq3)$, and the corresponding TQt-SVD is $\dot{\mathcal{T}}= \dot{\mathcal{U}}\star_{QT}\dot{\mathcal{D}}\star_{QT}\dot{\mathcal{V}}^H$. The TQt-rank of $\dot{\mathcal{T}}$ is defined as  the number of nonzero tubes in $\dot{\mathcal{D}}$, \textit{i.e.}, $rank_{TQt}(\dot{\mathcal{T}})= \#\{k\mid\parallel\dot{\mathcal{D}}(k, k,:\cdots, :)\parallel_F>0\}, k\in[K], K=\min\{I_1,I_2\}$. Moreover, the \textit{k}th singular value of $\dot{\mathcal{T}}$ is defined as $\sigma_{k}(\dot{\mathcal{T}})=\parallel\dot{\mathcal{D}}(k, k,:\cdots, :)\parallel_F$.
\end{definition}

More detail basis about quaternions and quaternion tensors can be found in \cite{DBLP:journals/corr/abs-2112-08921, DBLP:journals/pr/MiaoKL20, zhang1997quaternions}.

\section{Quaternion Tensor Low-rank Approximation and Sparse Regularization}
\label{Main1}
\subsection{Low-rank Approximation for Quaternion Tensor}
\begin{figure} [htbp]
	\centering
	\includegraphics[width=90mm]{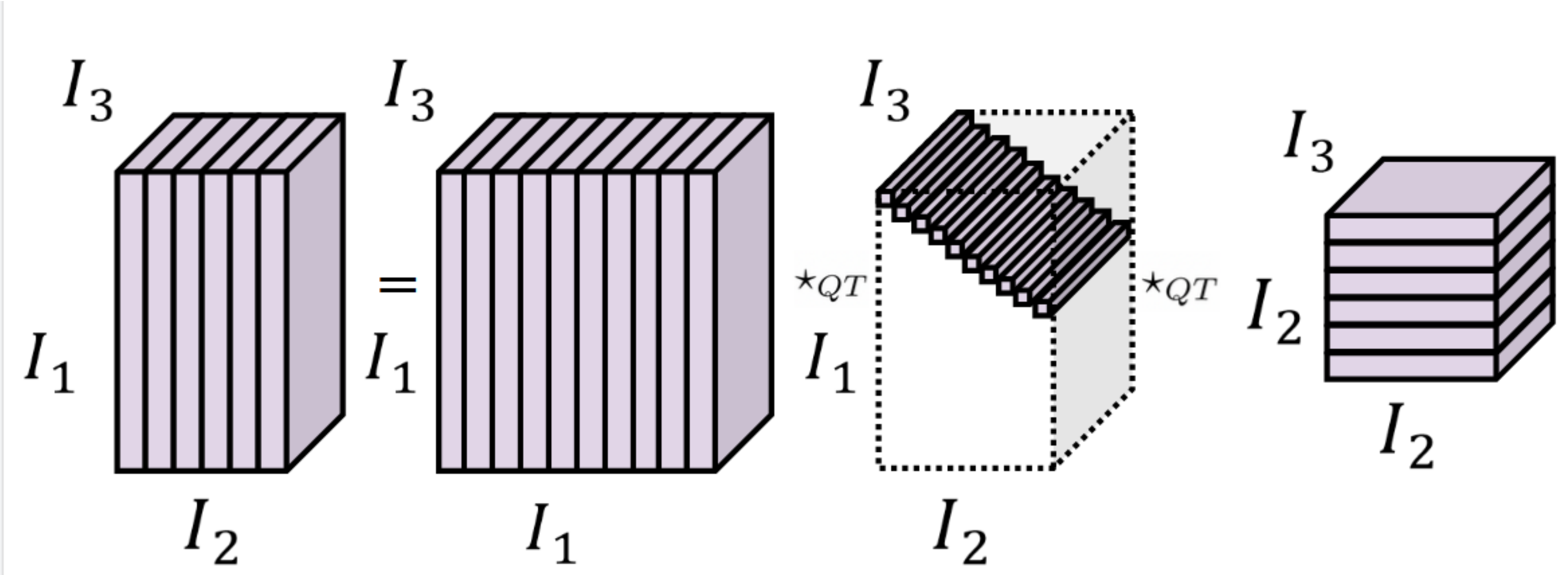}
	\caption{The illustration of the TQt-SVD  of an $I_1\times I_2\times I_3$ quaternion tensor.}
	\label{f2}
\end{figure}

The TQt-rank can actually be regarded as a generalization of the real valued tensor tubal rank. Fig. \ref{f2}. provides an intuitive illustration of the three-dimensional TQt-SVD. Additionally, based on the definition of $\star_{QT}$-product, the specific process of computing TQt-SVD of the target quaternion tensor is to perform QSVD for every frontal slice in the transformed domain that is equivalent to the invertible quaternion matrix $\dot{\mathbf{Q}}_3$ mode-product in the $\star_{QT}$-product. Let $\mathit{L}$ be the linear transform $\mathbb{H}^{I_1\times I_2\times I_3}\longrightarrow \mathbb{H}^{I_1\times I_2\times I_3}$: $\hat{\dot{\mathcal{T}}}=\mathit{L}(\dot{\mathcal{T}})=\dot{\mathcal{T}}\times_3\dot{\mathbf{Q}}_3$. Similarly, the inverse mapping can be formulated as $\mathit{L}^{-1}(\dot{\mathcal{T}})=\dot{\mathcal{T}}\times_3\dot{\mathbf{Q}}_3^{-1}$.

The particulars of the process of TQt-SVD can be seen in Algorithm \ref{a1}. 

\begin{algorithm}[htbp]
	\caption{TQt-SVD based on $\star_{QT}$-product}
	\label{a1}
	\begin{algorithmic}[1]
		\REQUIRE   $\dot{\mathcal{T}}\in\mathbb{H}^{I_1\times I_2\times\times I_3}$, and invertible quaternion matrix $\dot{\mathbf{Q}}_3\in\mathbb{H}^{I_3\times I_3}$. 
		\STATE Compute $\hat{\dot{\mathcal{T}}}=\mathit{L}(\dot{\mathcal{T}})=\dot{\mathcal{T}}\times_3\dot{\mathbf{Q}}_3$ 
		\STATE Compute each frontal slice of  $\hat{\dot{\mathcal{U}}}$, $\hat{\dot{\mathcal{D}}}$, $\hat{\dot{\mathcal{V}}}$ from $\hat{\dot{\mathcal{T}}}$ by\\
			\textbf{for} $i=1,\cdots, n_3$ \textbf{do}\\
			$[\hat{\dot{\mathcal{U}}}^{(i)}, \hat{\dot{\mathcal{D}}}^{(i)}, \hat{\dot{\mathcal{V}}}^{(i)}]=QSVD(\hat{\dot{\mathcal{T}}}^{(i)})$;\\
		   \textbf{end for}
		\STATE Compute $\dot{\mathcal{U}}=\mathit{L}^{-1}(\hat{\dot{\mathcal{U}}}), \dot{\mathcal{D}}=\mathit{L}^{-1}(\hat{\dot{\mathcal{D}}}), \dot{\mathcal{V}}=\mathit{L}^{-1}(\hat{\dot{\mathcal{V}}}).$ 
		\ENSURE  TQt-SVD components $\dot{\mathcal{U}}, \dot{\mathcal{D}}$, and $\dot{\mathcal{V}}$ of $\dot{\mathcal{T}}$.
	\end{algorithmic}
\end{algorithm}

From Theorem \ref{th1}, it can be concluded that the diagonal matrix is real such that the f-diagonal quaternion tensor in TQt-SVD is a real quaternion tensor in the transform domain.  Moreover, it can be observed that the singular values of $\dot{\mathcal{T}}^{(i)}$ and $\dot{\mathcal{D}}^{(i)}$ are the same, whiche are the elements on the diagonal of $\dot{\mathcal{D}}^{(i)}$. Based on the above discussion, the quaternion tensor nuclear norm is defined for three-dimensional quaternion tensor as follows.

\begin{definition}[\textbf{quaternion tensor tubal rank and nuclear norm (QTNN)}] For  $\dot{\mathcal{T}}\in\mathbb{H}^{I_1\times I_2\times \times I_3}$, its corresponding TQt-SVD is $\dot{\mathcal{T}}= \dot{\mathcal{U}}\star_{QT}\dot{\mathcal{D}}\star_{QT}\dot{\mathcal{V}}^H$. The tubal rank of $\dot{\mathcal{T}}$ is defined as  the maximum rank among all frontal slices of the f-diagonal quaternion tensor $\hat{\dot{\mathcal{D}}}$ \textit{i.e.}, $rank_{TQt}(\dot{\mathcal{T}})= \max_i\{rank(\dot{\mathcal{D}}^{(i)})\}$. The quaternion tensor nuclear norm is defined as the sum of  singular values of   $\dot{\mathcal{T}}$, \textit{i.e.}, $\parallel\dot{\mathcal{T}}\parallel_*=\sum_{i=1}^{I_3}\sum_{j=1}^{\min\{I_1, I_2\}}\sigma_j(\dot{\mathcal{T}}^{(i)})=\sum_{i=1}^{I_3}\sum_{j=1}^{\min\{I_1, I_2\}}\sigma_j(\dot{\mathcal{D}}^{(i)})=tr(\dot{\mathcal{T}})$. 
\label{def6}
\end{definition}

As highlighted in \cite{DBLP:journals/pami/HuZYLH13}, the largest  singular values do not affect the rank of target matrix.Various works were inspired by it and subsequently improved the efficacy of image completion \cite{DBLP:conf/icpr/XueQLJ18, DBLP:journals/jvcir/YangKM21}. Furthermore, the quaternion tensor truncated nuclear norm is proposed as follows.
\begin{definition}[\textbf{quaternion tensor truncated nuclear norm (QT-RNN)}] For  $\dot{\mathcal{T}}\in\mathbb{H}^{I_1\times I_2\times \times I_3}$, the QT-RNN is defined as $\|\dot{\mathcal{T}}\|_r=\sum_{i=1}^{I_3}\sum_{j=r+1}^{\min\{I_1, I_2\}}\sigma_j(\dot{\mathcal{T}}^{(i)}).$
	\label{def7}
\end{definition}

Motivated by the superior performance of logarithmic function acts on singular values compared to NN \cite{kang2016top, DBLP:journals/tip/ChenJLZ21, ji2017non}, the quaternion tensor Logarithmic norm (QTLN) is defined as follows.
\begin{definition}[\textbf{QTLN}] Given $\dot{\mathcal{T}}\in\mathbb{H}^{I_1\times I_2\times \times I_3}$, the logarithmic norm of the quaternion tensor with $0<p \leq1$ and $\epsilon>0$ is:
	\begin{equation}
		\parallel\dot{\mathcal{T}}\parallel_L^p=\sum_{i=1}^{\min(M,N)}\log(\sigma_{i}^p(\dot{\mathcal{T}})+\epsilon),
	\end{equation}
	where $\sigma_{i}$  can be obtained by the TQt-SVD of  $ \dot{\mathcal{T}}$.
	\label{def8}
\end{definition}

The logarithmic function can flexibly control the penalization of singular values, which is consistent with the operation that larger singular values should be punished less (because the larger the singular value, the less it will affect the judgment of the rank) and correspondingly, smaller singular values should be punished more.

\subsection{Sparsity of The Underlying Quaternion Tensor}
Many nature signals are sparse, like audios, images, and videos, and such property can be characterized in a certain transformed domain \cite{wright2010sparse}. Motivated by these factors, in this section, the QTDCT is defined without destroying the structure of the quaternion tensor and the practical steps are listed.

\begin{definition}[\textbf{QTDCT}]  As the non-commutative multiplication of quaternions, QTDCT has two forms, 
left-handed form $\text{QTDCT}_L  (\mathcal{C}(\cdot)_L)$ and right-handed form $\text{QTDCT}_R (\mathcal{C}(\cdot)_R)$. They satisfy the following formulas, respectively. Let  $\dot{\mathcal{T}}\in\mathbb{H}^{I_1\times I_2\times\times I_3}$ be a input signal, the transformed output quaternion tensor $\mathcal{C}(\dot{\mathcal{T}})\in\mathbb{H}^{I_1\times I_2\times\times I_3}$ is given by:
	\begin{equation}
		\mathcal{C}(\dot{\mathcal{T}})_L\triangleq\dot{u}\cdot(\dot{\mathcal{T}}\times_1\mathbf{C}_1\times_2\mathbf{C}_2\times_3\mathbf{C}_3)
	\end{equation}
\begin{equation}
	\mathcal{C}(\dot{\mathcal{T}})_R\triangleq(\dot{\mathcal{T}}\times_1\mathbf{C}_1\times_2\mathbf{C}_2\times_3\mathbf{C}_3)\cdot\dot{u},
\end{equation}
where $\dot{u}$ is a pure quaternion and meets $\dot{u}^2=-1$, and $\mathbf{C}_i\in\mathbb{R}^{I_i\times I_i}, i=1, 2, 3.$ are DCT matrix as described in \cite{DBLP:journals/tip/CoutinhoCB17}.
\end{definition}

Corresponding to QTDCT, there are two forms of inverse transformation of QTDCT (denoted as $\mathcal{C}^{-1}(\cdot)_L$ and $\mathcal{C}^{-1}(\cdot)_R$). They satisfy the following formulas, respectively: 
\begin{equation}
	\mathcal{C}^{-1}(\dot{\mathcal{T}})_L\triangleq\dot{u}\cdot(\dot{\mathcal{T}}\times_1\mathbf{C}_1^{-1}\times_2\mathbf{C}_2^{-1}\times_3\mathbf{C}_3^{-1})
\end{equation}
\begin{equation}
	\mathcal{C}^{-1}(\dot{\mathcal{T}})_R\triangleq(\dot{\mathcal{T}}\times_1\mathbf{C}_1^{-1}\times_2\mathbf{C}_2^{-1}\times_3\mathbf{C}_3)^{-1}\cdot\dot{u},
\end{equation}
$\mathcal{C}(\dot{\mathcal{T}})$ and $\mathcal{C}^{-1}(\dot{\mathcal{T}})$ are the transformation pairs of each other.
	
\begin{figure*}[htbp]
	\captionsetup{font={footnotesize}}
	\subfigure{
		\begin{minipage}{6cm}
			\centering
			\includegraphics[height = 3.5cm, width = 4 cm]{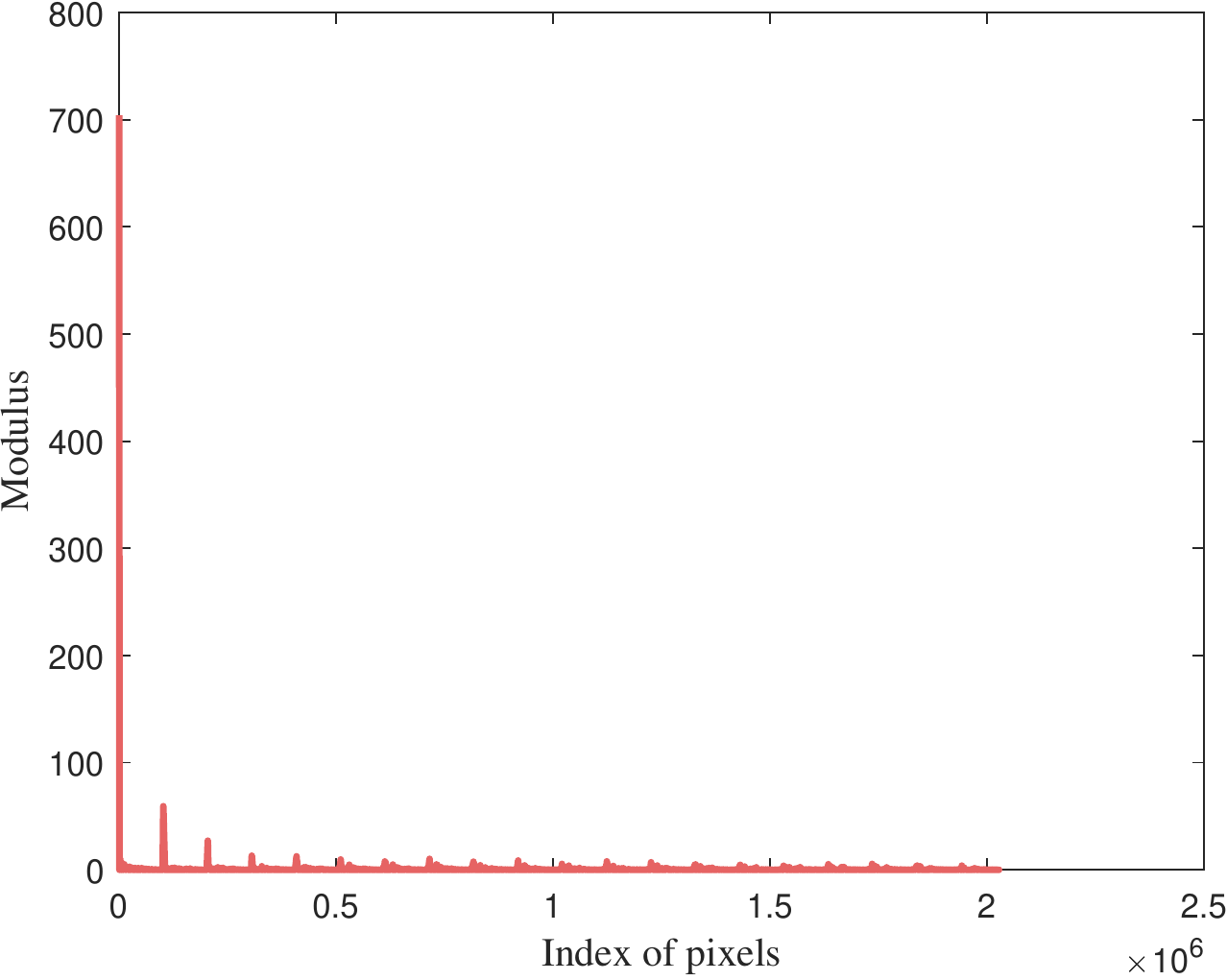}
		\end{minipage}
	}
	\subfigure{
		\begin{minipage}{6cm}
			\centering
			\includegraphics[height = 3.5cm, width = 4cm]{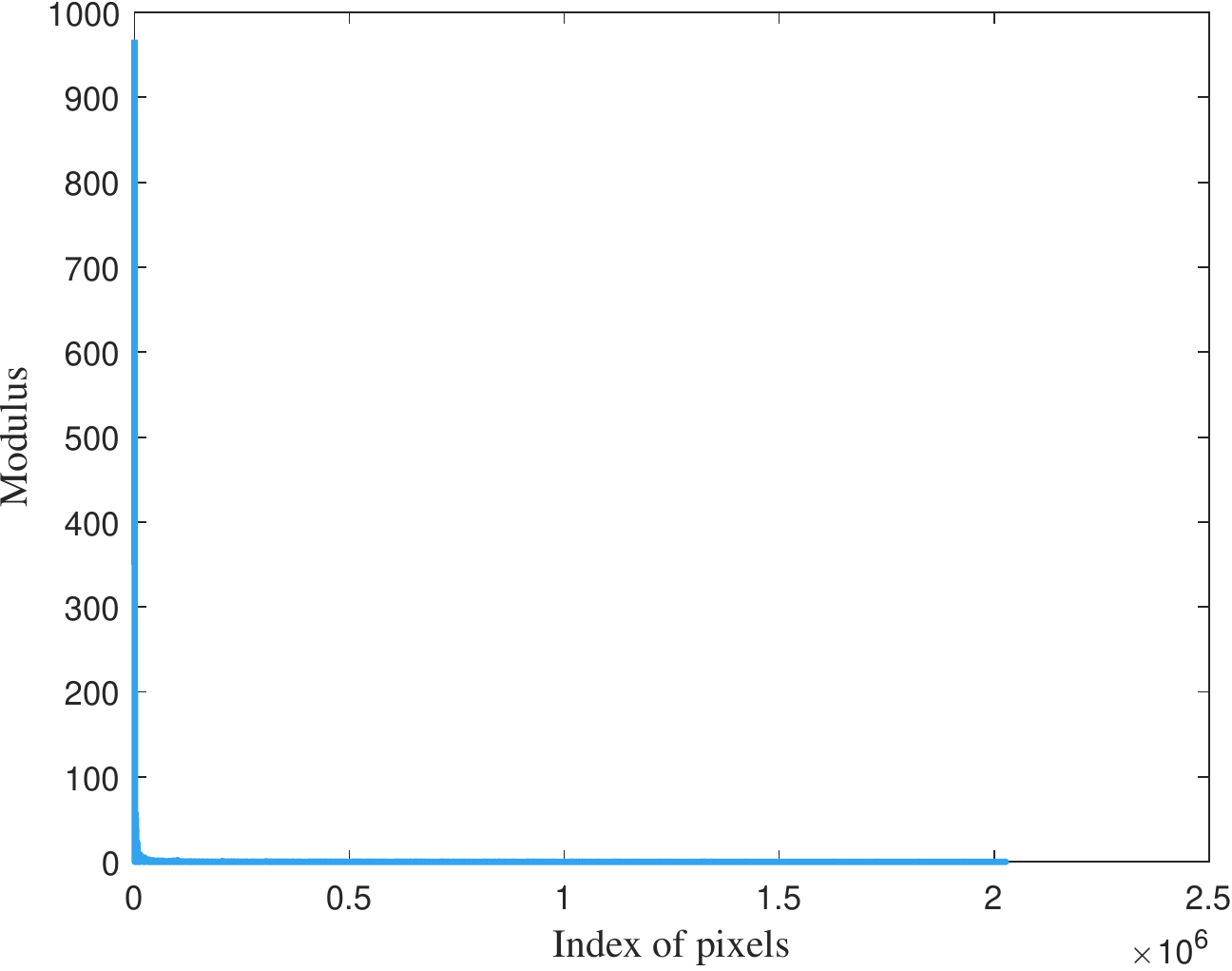}
		\end{minipage}
	}
	\subfigure{
		\begin{minipage}{6cm}
			\centering
			\includegraphics[height = 3.5cm, width = 4cm]{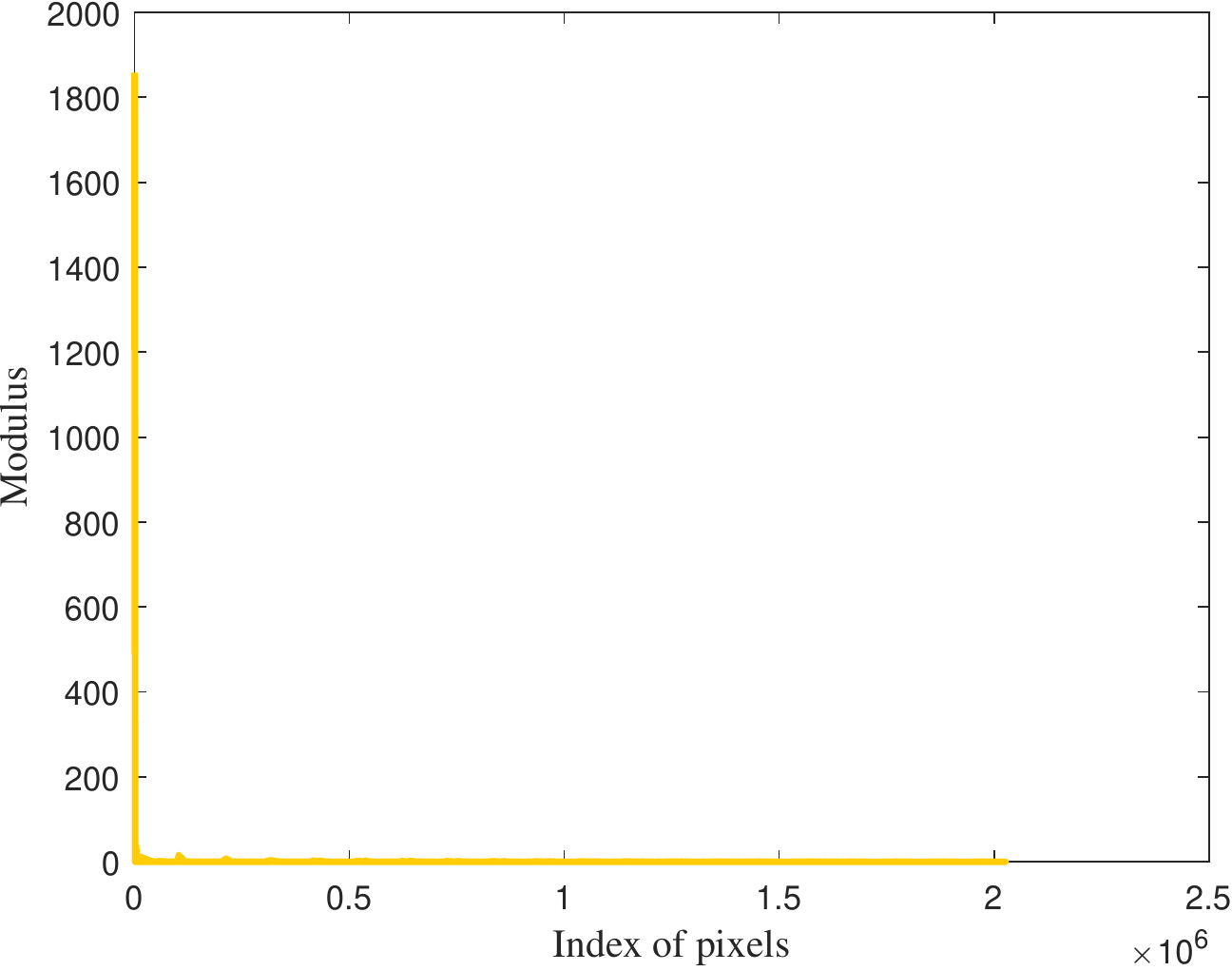}
		\end{minipage}
	}
	\caption{The illustration of the sparsity under QTDCT, taking three color videos (each video has 20 frames) as examples.}
	\label{f4}
\end{figure*}

In the developed algorithm, $\text{QTDCT}_L$ is utilized to carry out QTDCT. A better explanation of the efficiency of $\text{QTDCT}_L$ is that it is used to depict the sparsity of the target quaternion tensor. In Fig. \ref{f4}, the modulus of each pixel of the color video stream is drawn. Typically, the modulus of the pixels of the color video stream  tend zero after QTDCT transformation, which means that the video stream is sparse in the transform domain. This is the sparse regularization which is employed in the proposed algorithm.
Taking advantage of the Cayley Dickson notation and motivating by the matrix form in Definition \ref{def1}, all steps of $\text{QTDCT}_L$ are listed as follows:
	\begin{enumerate}
		\item[(a)] Transforming the given quaternion tensor $\dot{\mathcal{T}}\in\mathbb{H}^{I_1\times I_2\times\times I_3}$  to the Cayley Dickson form $\dot{\mathcal{T}}=\mathcal{T}_p+\mathcal{T}_qj$, where $\mathcal{T}_p$ and $\mathcal{T}_q\in\mathbb{C}^{I_1\times I_2\times\times I_3}$.
		\item[(b)] Calculating the multidimensional DCT of complex tensors $\mathcal{T}_p$ and $\mathcal{T}_q$. The results are denoted as $\text{DCT}_C(\mathcal{T}_p)$ and $\text{DCT}_C(\mathcal{T}_q)$, respectively.
		\item[(c)] According to Cayley Dickson theorem,  $\text{DCT}_C(\mathcal{T}_p)$ and $\text{DCT}_C(\mathcal{T}_q)$ can form a quaternion tensor: $	\hat{\mathcal{C}}(\dot{\mathcal{T}})_L=\text{DCT}_C(\mathcal{T}_p)+\text{DCT}_C(\mathcal{T}_q)j$.
		\item[(d)] Multiplying $\mathcal{C}(\dot{\mathcal{T}})_L$ with the unit quaternion factor $\dot{u}$ to obtain the final result $\mathcal{C}(\dot{\mathcal{T}})_L=\dot{u}\cdot\hat{\mathcal{C}}(\dot{\mathcal{T}})_L$.
	\end{enumerate}

\begin{figure*}[htbp]
	\captionsetup{font={footnotesize}}
	\subfigure{
		\begin{minipage}{6cm}
			\centering
			\includegraphics[height = 3.5cm, width = 4 cm]{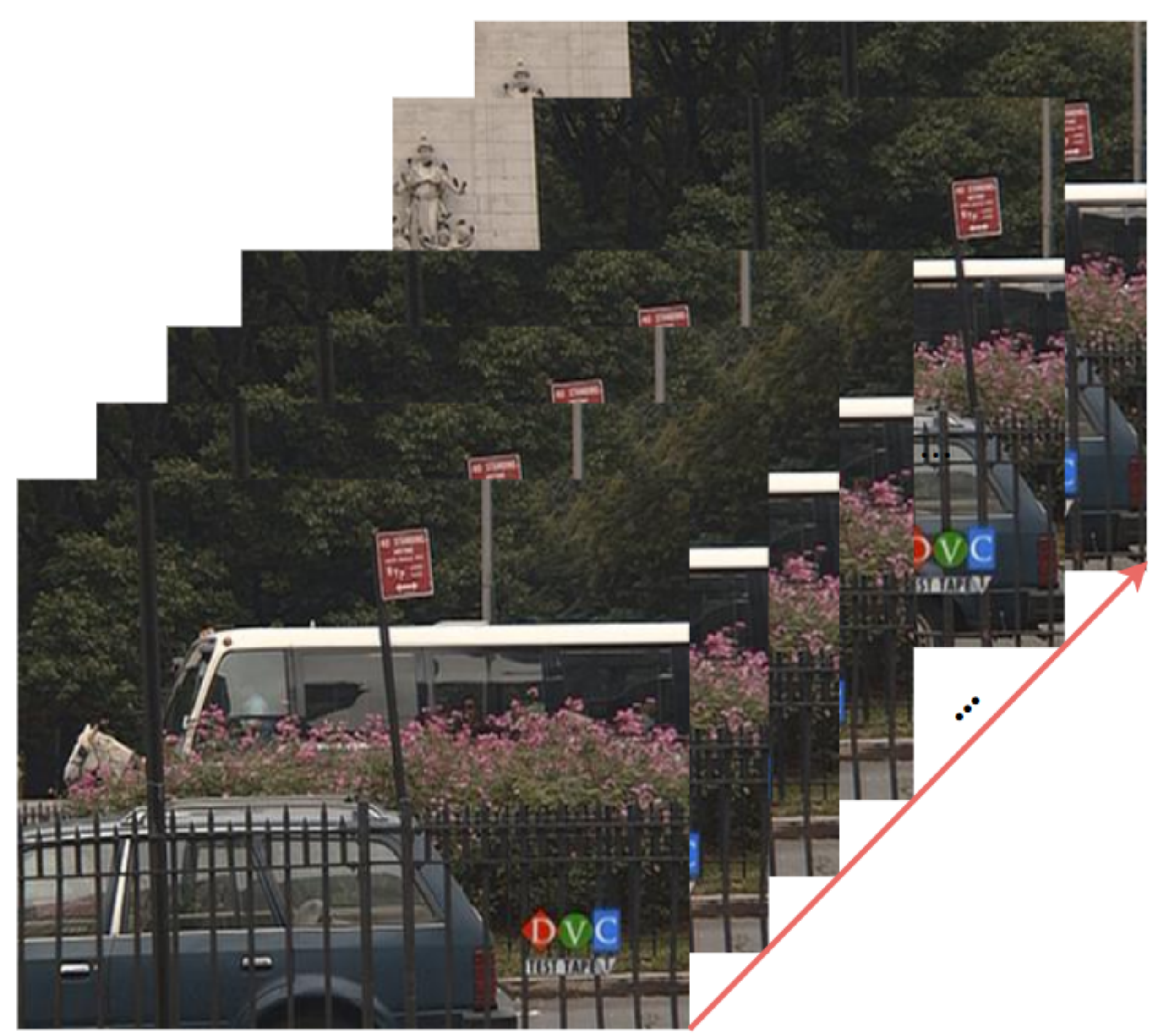}
		\end{minipage}
	}
	\subfigure{
		\begin{minipage}{6cm}
			\centering
			\includegraphics[height = 3.5cm, width = 4cm]{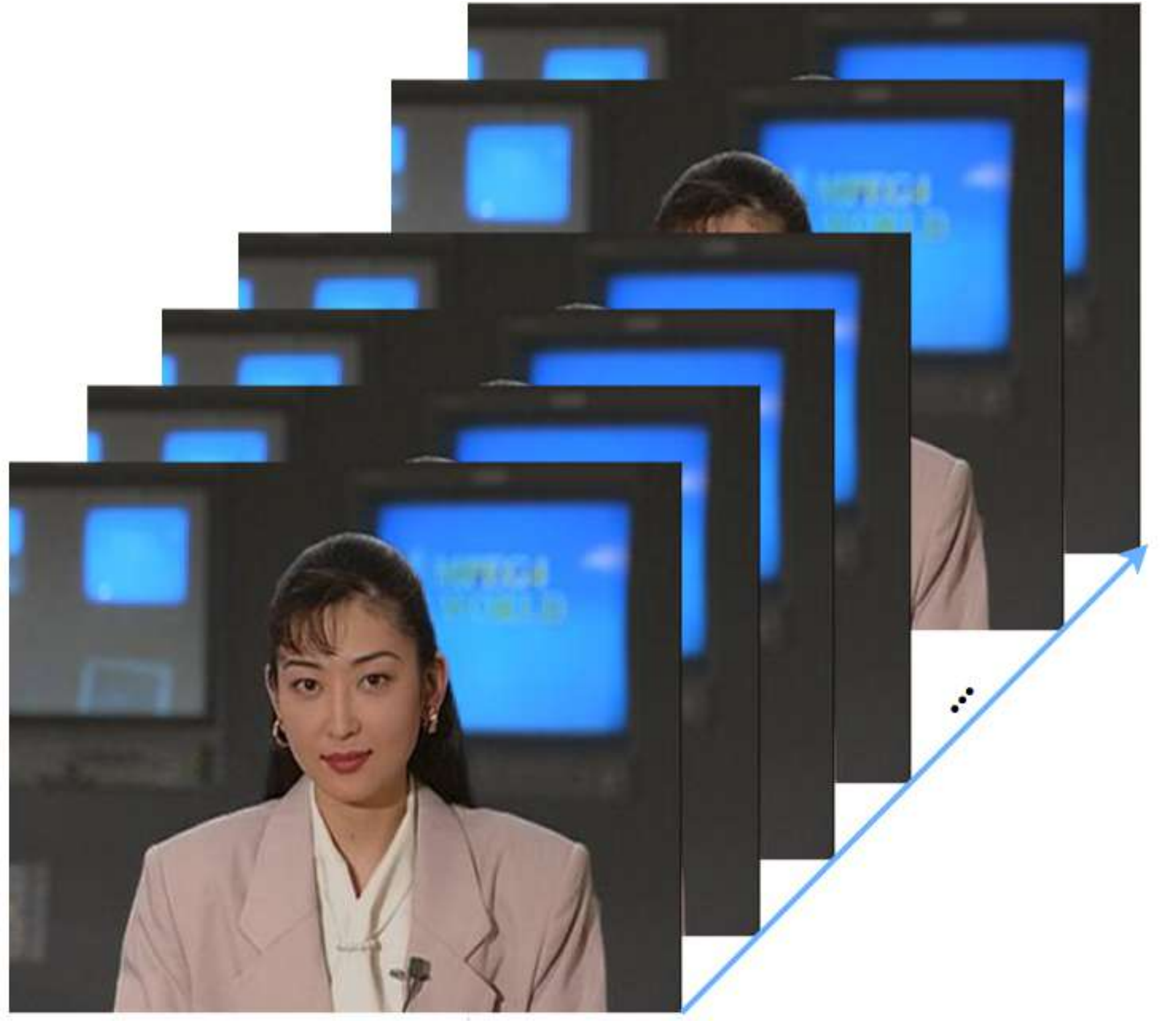}
		\end{minipage}
	}
	\subfigure{
	\begin{minipage}{6cm}
		\centering
		\includegraphics[height = 3.5cm, width = 4cm]{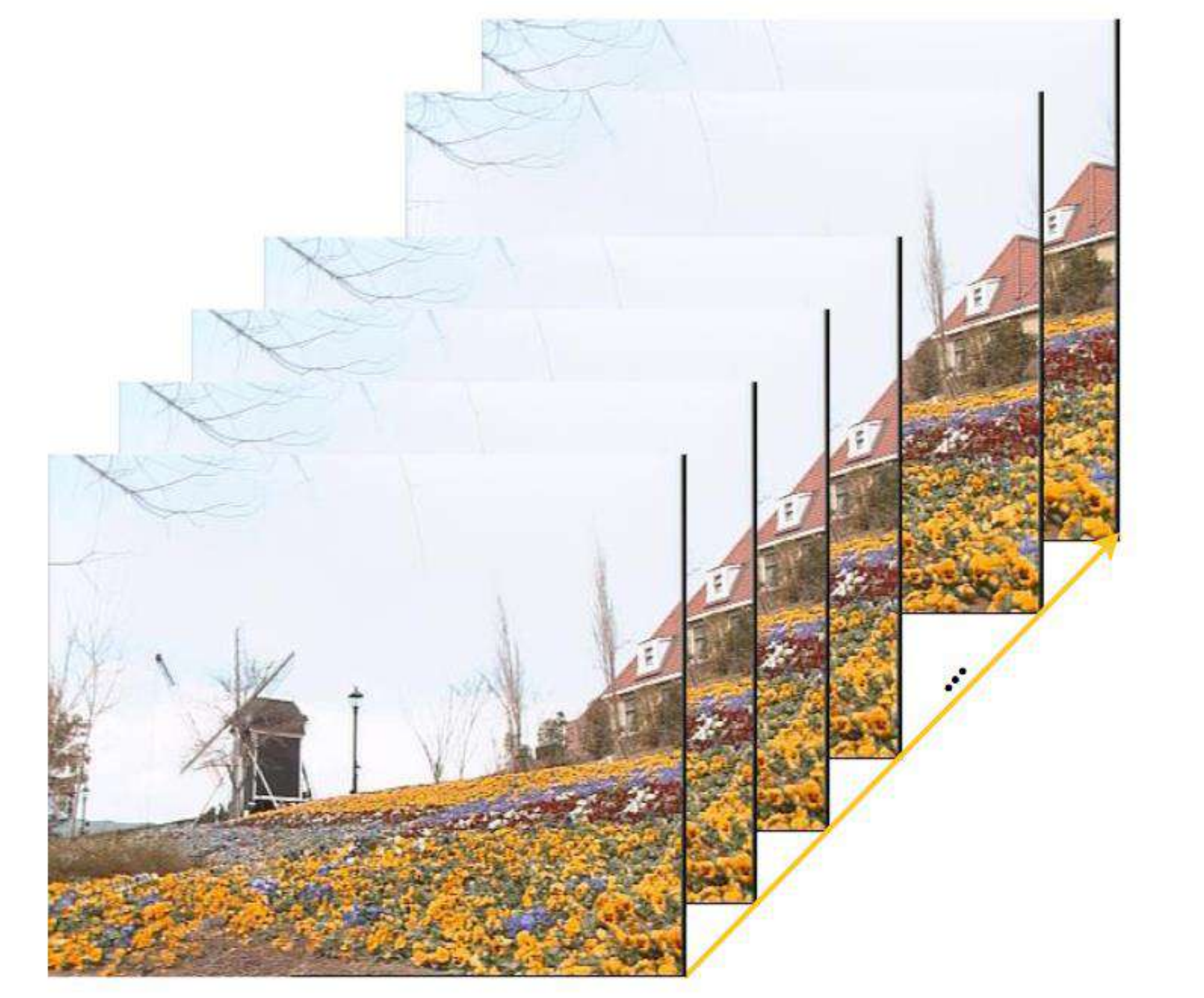}
	\end{minipage}
}
	\label{f3}
\end{figure*}

\section{ The corresponding recovery models and Algorithm}
\label{Main2}
This section is divided into two parts. Subsection \ref{3.1} gives the formulation of the proposed recovery model, while subsection \ref{3.2} presents the two-step optimization with ADMM.
\subsection{Problem Formulation}
\label{3.1}
This section gives the formulation of low-rank quaternion tensor optimized model for color video recovery. 

Similar to completion models based on low-rank constraints, thee low-rank quaternion tensor completion also utilizes low rankness as previously detailed to build the relationship between the observed and the missing elements in the target quaternion tensor. Mathematically, the recovery model can be formulated as follows.
 \begin{equation}\label{m1}
 	\begin{aligned}
 		&\min\limits_{\dot{\mathcal{T}}} \text{rank}(\dot{\mathcal{T}})\\& \text{ s.t.}   \quad   P_\Omega(\dot{\mathcal{T}})=P_\Omega(\dot{\mathcal{O}}),
 	\end{aligned}
 \end{equation}
where $\dot{\mathcal{T}}$ is the underlying quaternion tensor, $\dot{\mathcal{O}}$ is the observed quaternion tensor, $\Omega$ is the index set for observed elements. $P_\Omega$ is a linear operation that indicates the elements in $\Omega$ are unchanged and zeros out others. 

In the existing research, there has been extensive discussion of the nuclear norm (sum of the singular values) as the tightest convex substitute of rank. Moreover, the first few largest singular values do not affect the rank, hence, the QT-RNN which proposed in Definition \ref{def7} is adopted to replace the nuclear norm. Additionally, the sparsity as a vital property is considered in the recovery model by the designed regularization under QTDCT. Subsequently, the QT-RNN with sparsity (QT-RNNS) model can be written as follows.
\begin{equation}\label{m2}
	\begin{aligned}
		&\min\limits_{\dot{\mathcal{T}}}\parallel\dot{\mathcal{T}}\parallel_r+\lambda \parallel\dot{\mathcal{S}}\parallel_1\\&\text{s.t.}  \quad P_\Omega(\dot{\mathcal{T}})=P_\Omega(\dot{\mathcal{O}}), \dot{\mathcal{S}}=\mathcal{C}(\dot{\mathcal{T}})_L,
	\end{aligned}
\end{equation} 
where $\parallel\dot{\mathcal{T}}\parallel_r=\sum_{i=1}^{I_3}\sum_{j=r+1}^{\min\{I_1,I_2\}}\sigma_j(\dot{\mathcal{T}}^{(i)})=\sum_{i=1}^{I_3}\sum_{j=1}^{\min\{I_1,I_2\}}\sigma_j(\dot{\mathcal{T}}^{(i)})-\sum_{i=1}^{I_3}\sum_{j=1}^{r}\sigma_j(\dot{\mathcal{T}}^{(i)})=\parallel\dot{\mathcal{T}}\parallel_*-\sum_{i=1}^{I_3}\sum_{j=1}^{r}\sigma_j(\dot{\mathcal{T}}^{(i)})$, $\dot{\mathcal{S}}$ is the transformed quaternion tensor,  $\mathcal{C}(\cdot)$ is the QTDCT transform operator, and $\lambda$ is a positive number. 

Basing on the Theorem 3 in \cite{DBLP:journals/jvcir/YangKM21}, the equation in model (\ref{m2}) can be rewritten as 
\begin{equation}\label{m3}
	\begin{aligned}
		&\min\limits_{\dot{\mathcal{T}}}\parallel\dot{\mathcal{T}}\parallel_*- \sum_{i=1}^{I_3}\mathop{\max}\limits_{\dot{\mathbf{A}}^{(i)}\dot{\mathbf{A}}^{(i)H}=\mathbf{I},\atop \dot{\mathbf{B}}^{(i)}\dot{\mathbf{B}}^{(i)H}=\mathbf{I}} |tr(\dot{\mathbf{A}}^{(i)}\dot{\mathcal{T}}^{(i)}\dot{\mathbf{B}}^{(i)H})|+\lambda \parallel\dot{\mathcal{S}}\parallel_1\\&\text{s.t.}  \quad P_\Omega(\dot{\mathcal{T}})=P_\Omega(\dot{\mathcal{O}}), \dot{\mathcal{S}}=\mathcal{C}(\dot{\mathcal{T}})_L,
	\end{aligned}
\end{equation} 
where $\dot{\mathbf{A}}^{(i)}$ and $\dot{\mathbf{B}}^{(i)}$ are combined by the first r columns of  $\dot{\mathcal{U}}^{(i)}$ and $\dot{\mathcal{V}}^{(i)}$ respectively, which are the left and right unitary matrices of $\dot{\mathcal{T}}^{(i)}$. Besides, based on the definition of $\star_{QT}$-product in Definition \ref{def4}, \eqref{m3} can be converted to
\begin{equation}\label{m4}
	\begin{aligned}
		&\min\limits_{\dot{\mathcal{T}}}\parallel\dot{\mathcal{T}}\parallel_*-\mathop{\max}\limits_{\dot{\mathcal{A}}\star_{QT}\dot{\mathcal{A}}^{H}=\dot{\mathcal{I}},\atop \dot{\mathcal{B}}\star_{QT}\dot{\mathcal{B}}^{H}=\dot{\mathcal{I}}} |tr(\dot{\mathcal{A}}\star_{QT}\hat{\dot{\mathcal{T}}}\star_{QT}\dot{\mathcal{B}}^{H})|+\lambda \parallel\dot{\mathcal{S}}\parallel_1\\&\text{s.t.}  \quad P_\Omega(\dot{\mathcal{T}})=P_\Omega(\dot{\mathcal{O}}), \dot{\mathcal{S}}=\mathcal{C}(\dot{\mathcal{T}})_L,
	\end{aligned}
\end{equation} 
where $\dot{\mathcal{I}}$ is the identity quaternion tensor whose each frontal slice is the quaternion matrix. $\dot{\mathcal{A}}$ and $\dot{\mathcal{B}}$ is derived from the first $r$ columns of the second dimension of $\dot{\mathcal{U}}$ and $\dot{\mathcal{V}}$, respectively. Using Matlab notation, $\dot{\mathcal{A}}$ and $\dot{\mathcal{B}}$ can be denoted as follows
\begin{equation}
\dot{\mathcal{A}}\triangleq\dot{\mathcal{U}}(:,1:r,:)^H, \quad
\dot{\mathcal{B}}\triangleq\dot{\mathcal{V}}(:,1:r,:)^H.
\end{equation}

Thus, after obtaining $\dot{\mathcal{A}}$ and $\dot{\mathcal{B}}$ in the first step, the main purpose is to solve \eqref{m4} in the second step. The two-step based method is summarized in Algorithm \ref{a2}.

\begin{algorithm}[htbp]
	\caption{Low-rank Quaternion Recovery with Sparse Regularization}
	\label{a2}
	\begin{algorithmic}[1]
		\REQUIRE   the observed incomplete quaternion tensor $\dot{\mathcal{O}}\in\mathbb{H}^{I_1\times I_2\times I_3}$; the index set of observed elements $\Omega$; the tolerance $\varepsilon_0$.
		\STATE \textbf{Initialize} the initial number of iteration $l=1$, $\dot{\mathcal{T}}_1=\dot{\mathcal{O}}$.
		\STATE \textbf{Repeat}
		\STATE \quad \textbf{Step 1.} Calculating the QTt-SVD of the given $\dot{\mathbf{T}}_{l}$\\
		\qquad  \qquad \qquad  $[\dot{\mathcal{U}}_{l},\mathcal{D}_{l},\dot{\mathcal{V}}_{l}]=\text{QTt-SVD }(\dot{\mathcal{T}}_{l})$,\\
		 \quad where $\dot{\mathcal{U}}_{l}\in\mathbb{H}^{I_1\times I_1 \times I_3}$,
		$\dot{\mathcal{V}}_{l}\in\mathbb{H}^{I_2\times I_2 \times I_3}$ are the unitary \\ \quad quaternion tensor.
		\STATE   \quad Calculating $\dot{\mathcal{A}}_{l}$ and $\dot{\mathcal{B}}_{l}$ as follows\\ \quad
		$\dot{\mathcal{A}}_{l}\triangleq\dot{\mathcal{U}}_{l}(:,1:r,:)^H,  
		\dot{\mathcal{B}}_{l}\triangleq\dot{\mathcal{V}}_{l}(:,1:r,:)^H.$
		
		\STATE \quad \textbf{Step 2.} Solving the optimization problem as follows \\
		\qquad $\dot{\mathcal{T}}_{l+1}=\arg\min\limits_{\dot{\mathcal{T}}}\parallel\dot{\mathcal{T}}\parallel_*-|tr(\dot{\mathcal{A}}_l\star_{QT}\dot{\mathcal{T}}\star_{QT}\dot{\mathcal{B}}_l^H)| $\\\qquad \qquad\quad
		$+\lambda \parallel\dot{\mathcal{S}}\parallel_1$,\\
		\qquad  $\text{s.t.}  \quad P_\Omega(\dot{\mathcal{T}})=P_\Omega(\dot{\mathcal{O}}) \qquad \mathcal{C}(\dot{\mathcal{T}})=\dot{\mathcal{S}}.$
		
		\STATE \textbf{Until convergence} $\|\dot{\mathcal{T}}_{l+1}-\dot{\mathcal{T}}_{l}\|_F \leq \varepsilon_0$, $\dot{\mathcal{T}}_{opt}=\dot{\mathcal{T}}_{l+1}$.
		\ENSURE  the recovered quaternion tensor $\dot{\mathcal{T}}_{opt}$.
	\end{algorithmic}
\end{algorithm}
\subsection{ADMM-based optimization}
\label{3.2} 
Next, the goal is to optimize problem \eqref{m4}. The ADMM framework is employed to solve this problem. Firstly, an auxiliary variable is introduced $\dot{\mathcal{H}}$  and reformulating \eqref{m4} as 
\begin{equation}\label{m5}
	\begin{aligned}
		&\min\limits_{\dot{\mathcal{T}}}\parallel\dot{\mathcal{T}}\parallel_*-\mathop{\max}\limits_{\dot{\mathcal{A}}\star_{QT}\dot{\mathcal{A}}^{H}=\dot{\mathcal{I}},\atop \dot{\mathcal{B}}\star_{QT}\dot{\mathcal{B}}^{H}=\dot{\mathcal{I}}} |tr(\dot{\mathcal{A}}\star_{QT}\hat{\dot{\mathcal{H}}}\star_{QT}\dot{\mathcal{B}}^{H})|+\lambda \parallel\dot{\mathcal{S}}\parallel_1\\&\text{s.t.}  \quad P_\Omega(\dot{\mathcal{T}})=P_\Omega(\dot{\mathcal{O}}),
		\dot{\mathcal{H}}=\dot{\mathcal{T}}, 
		 \dot{\mathcal{S}}=\mathcal{C}(\dot{\mathcal{T}})_L.
	\end{aligned}
\end{equation} 
Inspiring by the ADMM framework utilized in the quaternion domain \cite{DBLP:journals/tsp/MiaoK20}. The corresponding augmented Lagrangian function of  \eqref{m5} can be written as 
\begin{equation}\label{m6}
	\begin{split}
		&\mathcal{L}(\dot{\mathcal{T}},\dot{\mathcal{H}},\dot{\mathcal{S}},\dot{\mathcal{Y}},\dot{\mathcal{Z}}, \beta)=\\&\parallel\dot{\mathcal{T}}\parallel_*- |tr(\dot{\mathcal{A}}_l\star_{QT}\dot{\mathcal{H}}\star_{QT}\dot{\mathcal{B}}_l^H)|+\lambda \parallel\dot{\mathcal{S}}\parallel_1\\&+\mathfrak{R}(\langle \dot{\mathcal{Y}}, \dot{\mathcal{T}}-\dot{\mathcal{H}} \rangle)+\frac{\beta}{2}\parallel\dot{\mathcal{T}}-\dot{\mathcal{H}}\parallel_F^2\\&+\mathfrak{R}(\langle\dot{\mathcal{Z}}, \dot{\mathcal{S}}-\mathcal{C}(\dot{\mathcal{T}})_L)
		+\frac{\beta}{2}\parallel\dot{\mathcal{S}}-\mathcal{C}(\dot{\mathcal{T}})_L\parallel_F^2,
	\end{split}
\end{equation}
where $\dot{\mathcal{Y}}$ and  $\dot{\mathcal{Z}}$ are Lagrange multipliers, and $\beta$ is a positive penalty parameter. With utilization of  ADMM algorithm, in the $k$-th iteration, \eqref{m6} can be rewritten as
\begin{equation}
	\left\{
	\begin{array}{lr}
		\dot{\mathcal{T}}^{k+1}=\arg\min\limits_{\dot{\mathcal{T}}}\mathcal{L}(\dot{\mathcal{T}},\dot{\mathcal{H}}^{k},\dot{\mathcal{S}}^{k},\dot{\mathcal{Y}}^{k},\dot{\mathcal{Z}}^{k}, \beta^k), &  \\
		\dot{\mathcal{S}}^{k+1}=\arg\min\limits_{\dot{\mathcal{S}}}\mathcal{L}(\dot{\mathcal{T}}^{k+1},\dot{\mathcal{H}}^{k},\dot{\mathcal{S}},\dot{\mathcal{Y}}^{k},\dot{\mathcal{Z}}^{k}, \beta^k),\\
		\dot{\mathcal{H}}^{k+1}=\arg\min\limits_{\dot{\mathcal{H}}}\mathcal{L}(\dot{\mathcal{T}}^{k+1},\dot{\mathcal{H}}, \dot{\mathcal{S}}^{k+1},\dot{\mathcal{Y}}^{k},\dot{\mathcal{Z}}^{k}, \beta^k),\\
		\dot{\mathcal{Y}}^{k+1}=\dot{\mathcal{Y}}^{k}+\beta^k(\dot{\mathcal{T}}^{k+1}-\dot{\mathcal{H}}^{k+1}),\\
		\dot{\mathcal{Z}}^{k+1}=\dot{\mathcal{Z}}^{k}+\beta^k(\dot{\mathcal{S}}^{k+1}-\mathcal{C}(\dot{\mathcal{T}}^{k+1})_L).
	\end{array}
	\right.
\end{equation}

\textbf{1. }The $\dot{\mathcal{T}}^{k+1}$ subproblem is
\begin{equation}\label{m7}
	\begin{split}
		\dot{\mathcal{T}}^{k+1}&=\arg\min\limits_{\dot{\mathcal{T}}}\parallel\dot{\mathcal{T}}\parallel_*+\mathfrak{R}(\langle \dot{\mathcal{Y}}^k, \dot{\mathcal{T}}-\dot{\mathcal{H}}^k \rangle)+\frac{\beta^k}{2}\parallel\dot{\mathcal{T}}-\dot{\mathcal{H}}^k\parallel_F^2\\&\quad+\mathfrak{R}(\langle\dot{\mathcal{Z}}^k, \dot{\mathcal{S}}^k-\mathcal{C}(\dot{\mathcal{T}}^k)_L)
		+\frac{\beta^k}{2}\parallel\dot{\mathcal{S}}^k-\mathcal{C}(\dot{\mathcal{T}})_L\parallel_F^2
		\\&=\arg\min\limits_{\dot{\mathcal{T}}}\parallel\dot{\mathcal{T}}\parallel_*+\frac{\beta^k}{2}\parallel\dot{\mathcal{T}}-\dot{\mathcal{H}}^k+\dot{\mathcal{Y}}^k/\beta^k\parallel_F^2\\&\quad+\frac{\beta^k}{2}\parallel\dot{\mathcal{S}}^k-\mathcal{C}(\dot{\mathcal{T}})_L+\dot{\mathcal{Z}}^k/\beta^k\parallel_F^2.
	\end{split}
\end{equation}
Although, in the last term of  \eqref{m7}, $\dot{\mathcal{T}}$ can not be separated directly since the transformation $\mathcal{C}$. The Parseval theorem is established in the quaternion domain, which demonstrates that after transforming the total energy of the signal is unchanged \cite{DBLP:journals/cma/BahriHHA08, DBLP:journals/corr/Hitzer13d}. This allows the last term to be computed in either the spatial domain or the quaternion domain. Thus the Frobenius norm in the last term of  \eqref{m7} can be rewritten as
\begin{equation*}
	\begin{split}
		\parallel\dot{\mathcal{S}}^k-\mathcal{C}(\dot{\mathcal{T}})_L+\dot{\mathcal{Z}}^k/\beta^k\parallel_F^2=\parallel\mathcal{C}^{-1}(\dot{\mathcal{S}}^k+\dot{\mathcal{Z}}^k/\beta^k)_L-\dot{\mathcal{T}}\parallel_F^2.
	\end{split}
\end{equation*}
Consequently, \eqref{m7} can be rewritten as
\begin{equation}\label{m8}
	\begin{split}
		\dot{\mathcal{T}}^{k+1}&=\arg\min\limits_{\dot{\mathcal{T}}}\parallel\dot{\mathcal{T}}\parallel_*+\frac{\beta^k}{2}\parallel\dot{\mathcal{T}}-\dot{\mathcal{H}}^k+\dot{\mathcal{Y}}^k/\beta^k\parallel_F^2\\&\quad+\frac{\beta^k}{2}\parallel\mathcal{C}^{-1}(\dot{\mathcal{S}}^k+\dot{\mathcal{Z}}^k/\beta^k)_L-\dot{\mathcal{T}}\parallel_F^2\\&=\arg\min\limits_{\dot{\mathcal{T}}}\parallel\dot{\mathcal{T}}\parallel_*\\&\quad+\beta^k\parallel\dot{\mathcal{T}}-\frac{1}{2}[\dot{\mathcal{H}}^k+\dot{\mathcal{Y}}^k/\beta^p+\mathcal{C}^{-1}(\dot{\mathcal{S}}^k+\dot{\mathcal{Z}}^k/\beta^k)]\parallel_F^2.
	\end{split}
\end{equation}

To solve \eqref{m8}, the Quaternion Tensor Singular Value Thresholding (QTSVT) is defined as follows
\begin{definition}[QTSVT] Let $\dot{\mathcal{X}}= \dot{\mathcal{U}}\star_{QT}\dot{\mathcal{D}}\star_{QT}\dot{\mathcal{V}}^H$ be the QTt-SVD of $\dot{\mathcal{X}}$. For any $\tau>0$, the QTSVT operator is defined as
	\begin{equation}\label{m9}
\mathfrak{D}_{\tau}(\dot{\mathcal{X}})=\dot{\mathcal{U}}\star_{QT}\dot{\mathcal{D}}_{\tau}\star_{QT}\dot{\mathcal{V}}^H, 
	\end{equation}
where  $\dot{\mathcal{D}}_{\tau}=L^{-1}((L(\dot{\mathcal{D}})-\tau))_+$, and $(*)_+$ denotes the positive part of $*$.  
\end{definition}
\begin{theorem}\label{th3}
Let $L$ be the linear transform which is utilized in TQt-SVD, for any $\tau>0$ and $\dot{\mathcal{X}}\in\mathbb{H}^{I_1\times I_2\times I_3}$, the TQSVT in \eqref{m9} obeys
\begin{equation}
\mathfrak{D}_{\tau}(\dot{\mathcal{X}})=\arg\min\limits_{\dot{\mathcal{T}}}\tau\parallel\dot{\mathcal{T}}\parallel_*+\frac{1}{2}\parallel\dot{\mathcal{T}}-\dot{\mathcal{X}}\parallel_F^2.
\end{equation}
\end{theorem}

Utilizing Theorem \ref{th3}, \eqref{m8} can be solved by
\begin{equation}\label{m10}
	\dot{\mathcal{T}}^{k+1}= \mathfrak{D}_{\frac{1}{2\beta^k}}(\frac{1}{2}[\dot{\mathcal{H}}^k+\dot{\mathcal{Y}}^k/\beta^k+\mathcal{C}^{-1}(\dot{\mathcal{S}}^k+\dot{\mathcal{Z}}^k/\beta^k)]).
\end{equation}

It can be observed from \eqref{m8} that the process of obtaining the update of $\dot{\mathcal{T}}^{k+1}$ is to find the optimized solution from the sum of QTNN and Frobenius norm. In this process, the QTNN also implies that all the singular values are treated equally, to optimize this problem and get a more precise approximation of the rank, the QTNN is further replaced by QTLN in \eqref{m8}. In this way, \eqref{m8} can be reformulated as

\begin{equation}\label{m81}
	\begin{split}
		\dot{\mathcal{T}}^{k+1}&=\arg\min\limits_{\dot{\mathcal{T}}}\parallel\dot{\mathcal{T}}\parallel_L+\frac{\beta^k}{2}\parallel\dot{\mathcal{T}}-\dot{\mathcal{H}}^k+\dot{\mathcal{Y}}^k/\beta^k\parallel_F^2\\&\quad+\frac{\beta^k}{2}\parallel\mathcal{C}^{-1}(\dot{\mathcal{S}}^k+\dot{\mathcal{Z}}^k/\beta^k)_L-\dot{\mathcal{T}}\parallel_F^2\\&=\arg\min\limits_{\dot{\mathcal{T}}}\parallel\dot{\mathcal{T}}\parallel_L\\&\quad+\beta^k\parallel\dot{\mathcal{T}}-\frac{1}{2}[\dot{\mathcal{H}}^k+\dot{\mathcal{Y}}^k/\beta^p+\mathcal{C}^{-1}(\dot{\mathcal{S}}^k+\dot{\mathcal{Z}}^k/\beta^k)]\parallel_F^2.
	\end{split}
\end{equation}

For solving \eqref{m81}, similar to the real cases \cite{DBLP:journals/tip/ChenJLZ21}, the theorem of Quaternion Tensor Logarithmic Singular Value Thresholding (QTLSVT) is given as follows
\begin{theorem} [\textbf{QTLSVT}]
	\label{th4}
	For any quaternion tensor $\dot{\mathcal{X}}\in\mathbb{H}^{I_1\times I_2\times I_3}$ and $\lambda>0$,  the QTt-SVDt is $\dot{\mathcal{X}}= \dot{\mathcal{U}}\star_{QT}\dot{\mathcal{D}}\star_{QT}\dot{\mathcal{V}}^H$. Then the closed solution of the following problem
\begin{equation}
	\mathop{\arg\min}\limits_{\dot{\mathcal{T}}}\lambda\parallel\dot{\mathcal{T}}\parallel_L^1+ \frac{1}{2}\parallel \dot{\mathcal{T}} -\dot{\mathcal{X}}\parallel_F^2
\end{equation}
is provided by $\dot{\mathcal{U}}\star_{QT}\dot{\mathcal{L}}_{\lambda,\epsilon}\star_{QT}\dot{\mathcal{V}}^H$,
where  $\dot{\mathcal{L}}_{\lambda,\epsilon}(\dot{\mathcal{D}}_{\dot{\mathcal{X}}})=L^{-1}(\mathcal{L}_{\lambda,\epsilon}(L(\dot{\mathcal{D}})))$. The soft thresholding operator $\mathcal{L}_{\lambda, \epsilon}(\cdot)$ is defined as:
\begin{equation}\label{tm1}
	\mathcal{L}_{\lambda, \epsilon}(x)  =
	\begin{cases}
		0,\qquad \qquad \qquad  \qquad  \qquad \Delta \leq 0\\
	\mathop{\arg\min}\limits_{a\in \{0, \frac{1}{2}(x-\epsilon+\sqrt{\Delta})\}} \textit{h}(a), \qquad\Delta > 0,
	\end{cases}
\end{equation}
where $\Delta = (x-\epsilon)^2-4(\lambda-x\epsilon)$, and function $\mathit{h}(a) :=\frac{1}{2}(a-x)^2 + \lambda\log(a+\epsilon)$ is  $\mathbb{R}^+\longrightarrow\mathbb{R}^+$.
\end{theorem}

Utilizing Theorem \ref{th4}, \eqref{m81} can be solved by
\begin{equation}\label{m101}
	\dot{\mathcal{T}}^{k+1}= \dot{\mathcal{U}}\star_{QT}\dot{\mathcal{L}}_{\lambda,\epsilon}(\dot{\mathcal{D}}_{\dot{\mathcal{A}}})\star_{QT}\dot{\mathcal{V}}^H,
\end{equation}
where $\dot{\mathcal{A}}^k=\frac{1}{2}[\dot{\mathcal{H}}^k+\dot{\mathcal{Y}}^k/\beta^k+\mathcal{C}^{-1}(\dot{\mathcal{S}}^k+\dot{\mathcal{Z}}^k/\beta^k)]$.

\textbf{2.} The $\dot{\mathcal{S}}^{k+1}$ subproblem is 
\begin{equation}\label{m11}
	\begin{split}
		\dot{\mathcal{S}}^{k+1}&=\arg\min\limits_{\dot{\mathcal{S}}}\lambda\parallel\dot{\mathcal{S}}\parallel_1+\mathfrak{R}(\langle\dot{\mathcal{Z}}^k, \dot{\mathcal{S}}-\mathcal{C}(\dot{\mathcal{T}}^{k+1})_L)
		\\&\quad+\frac{\beta}{2}\parallel\dot{\mathcal{S}}-\mathcal{C}(\dot{\mathcal{T}}^{k+1})_L\parallel_F^2\\&
		=\arg\min\limits_{\dot{\mathcal{S}}}\lambda \parallel\dot{\mathcal{S}}\parallel_1+\frac{\beta^k}{2}\parallel\dot{\mathcal{S}}-\mathcal{C}(\dot{\mathcal{T}}^{k+1})+\dot{\mathcal{Z}}^{k}/\beta^k\parallel_F^2\\&
	\end{split}
\end{equation}

Because in the quaternion domain the closed solution of  $\min\limits_{\dot{\mathbf{X}}}\lambda \parallel\dot{\mathbf{X}}\parallel_1+\parallel\dot{\mathbf{Y}}-\dot{\mathbf{X}}\parallel_F^2$ is given by $ \dot{\mathbf{X}}_{opt}=\mathcal{S}_{2\lambda}(\dot{\mathbf{Y}})$ \cite{DBLP:journals/corr/abs-2204-08629}, analogously, problem \eqref{m11} has a closed-form solution given by 
\begin{equation}\label{m12}
	\dot{\mathcal{S}}^{k+1}=\mathcal{S}_{\frac{4\lambda}{\beta^k}}(\mathcal{C}(\dot{\mathcal{T}}^{k+1})-\dot{\mathcal{Z}}^{k}/\beta^k).
\end{equation}

\textbf{3. }The $\dot{\mathcal{H}}^{k+1}$ subproblem is 
\begin{equation}\label{m13}
	\begin{split}
		\dot{\mathcal{H}}^{k+1}&=\arg\min\limits_{\dot{\mathcal{H}}}- |tr(\dot{\mathcal{A}}_l\star_{QT}\dot{\mathcal{H}}\star_{QT}\dot{\mathcal{B}}_l^H)|\\&\quad+\mathfrak{R}(\langle \dot{\mathcal{Y}}^k, \dot{\mathcal{T}}^{k+1}-\dot{\mathcal{H}} \rangle)\\&\quad+\frac{\beta^k}{2}\parallel\dot{\mathcal{T}}^{k+1}-\dot{\mathcal{H}}\parallel_F^2.
	\end{split}
\end{equation}
Because \eqref{m13} is quadratic with respect to $\dot{\mathcal{H}}$, letting the derivation of \eqref{m13} be zero, then the closed solution of $\dot{\mathcal{H}}$ can be written as
\begin{equation}
	\dot{\mathcal{H}}^{k+1}=\dot{\mathcal{X}}^{k+1}+(\dot{\mathcal{Y}}^k+\dot{\mathcal{A}}^H\star_{QT}\dot{\mathcal{B}})/\beta^k.
\end{equation} 
Besides, let the observed elements remain unchanged in each iteration as follows
\begin{equation}
	\dot{\mathcal{H}}^{k+1}=P_{\Omega^C}(\dot{\mathcal{H}}^{k+1})+P_\Omega(\dot{\mathcal{O}}).
\end{equation} 

\textbf{4.} The update of Lagrange multiplier $\dot{\mathcal{Y}}^{k+1}$ 
\begin{equation}
\dot{\mathcal{Y}}^{k+1}=\dot{\mathcal{Y}}^{k}+\beta^k(\dot{\mathcal{T}}^{k+1}-\dot{\mathcal{H}}^{k+1}),\\
\end{equation}

\textbf{5.} The update of Lagrange multiplier $\dot{\mathcal{Z}}^{k+1}$ 
\begin{equation}
\dot{\mathcal{Z}}^{k+1}=\dot{\mathcal{Z}}^{k}+\beta^k(\dot{\mathcal{S}}^{k+1}-\mathcal{C}(\dot{\mathcal{T}}^{k+1})_L).
\end{equation}

\textbf{6.} The update of penalty parameter $\beta^k$ is
\begin{equation}
	\beta^{k+1}=\min\{\rho\beta^{k}, \beta_{\max}\},
\end{equation}
where $\rho \geq1$ is a constant parameter to increase the penalty, and $\beta_{\max}$ is the upper bounded of $\beta$.

The QTSVT-based method is denoted as QT-RNNS1, and the logarithmic QTLSVT-based method is denoted as QT-RNNS2. The whole process in \textbf{Step 2} is summarized in Algorithm \ref{a3}.
\begin{algorithm}[htbp]
	\caption{The process of QT-RNNS1 and QT-RNNS2 in \textbf{Step 2}.}
	\label{a3}
	\begin{algorithmic}[1]
		\REQUIRE    $\dot{\mathcal{O}}$, $\Omega$, $\dot{\mathcal{A}}_l$, $\dot{\mathcal{B}}_l$, tolerance $\varepsilon$, and parameters $\lambda$, $\rho$, $\beta_{\max}$.
		\STATE \textbf{Initial} $\dot{\mathcal{X}}^1=\dot{\mathcal{O}}$, $\dot{\mathcal{H}}^1=\dot{\mathcal{D}}^1=\dot{\mathcal{X}}^1$, and $\beta^1$. Let $\dot{\mathcal{Y}}^1$ and $\dot{\mathcal{Z}}^1$ be random quaternion matrix with the same size of $\dot{\mathcal{X}}^1$.
		\STATE \textbf{Repeat}
		\STATE Update $\dot{\mathcal{X}}^{k+1}$\\ $\dot{\mathcal{T}}^{k+1}= \mathfrak{D}_{\frac{1}{2\beta^k}}(\frac{1}{2}[\dot{\mathcal{H}}^k+\dot{\mathcal{Y}}^k/\beta^k+\mathcal{C}^{-1}(\dot{\mathcal{S}}^k+\dot{\mathcal{Z}}^k/\beta^k)]).$ \\Or\\
		$\dot{\mathcal{T}}^{k+1}=\dot{\mathcal{U}}\star_{QT}\dot{\mathcal{L}}_{\lambda,\epsilon}(\dot{\mathcal{D}}_{\dot{\mathcal{A}}})\star_{QT}\dot{\mathcal{V}}^H$,\\
where $\dot{\mathcal{A}}^k=\frac{1}{2}[\dot{\mathcal{H}}^k+\dot{\mathcal{Y}}^k/\beta^k+\mathcal{C}^{-1}(\dot{\mathcal{S}}^k+\dot{\mathcal{Z}}^k/\beta^k)]$.
		
		\STATE  Update $\dot{\mathcal{S}}^{k+1}=\mathcal{S}_{\frac{4\lambda}{\beta^k}}(\mathcal{C}(\dot{\mathcal{T}}^{k+1})-\dot{\mathcal{Z}}^{k}/\beta^k).$
		\STATE   Update $	\dot{\mathcal{H}}^{k+1}=\dot{\mathcal{X}}^{k+1}+(\dot{\mathcal{Y}}^k+\dot{\mathcal{A}}^H\star_{QT}\dot{\mathcal{B}})/\beta^k,$\\
		\qquad\quad$\dot{\mathcal{H}}^{k+1}=P_{\Omega^C}(\dot{\mathcal{H}}^{k+1})+P_\Omega(\dot{\mathcal{O}}).$
		\STATE Update $\dot{\mathcal{Y}}^{k+1}=\dot{\mathcal{Y}}^{k}+\beta^k(\dot{\mathcal{T}}^{k+1}-\dot{\mathcal{H}}^{k+1}).$
		\STATE Update
		$\dot{\mathcal{Z}}^{k+1}=\dot{\mathcal{Z}}^{k}+\beta^k(\dot{\mathcal{S}}^{k+1}-\mathcal{C}(\dot{\mathcal{T}}^{k+1})_L).$
		\STATE Update  $\beta^{k+1}=\min\{\rho\beta^{k}, \beta_{\max}\}$.
		\STATE \textbf{Until convergence} $\|\dot{\mathcal{T}}^{k+1}-\dot{\mathcal{T}}^{k}\|_F \leq \varepsilon$ or $p$ reaches the given maximum iteration  number. 
		\ENSURE  $\dot{\mathcal{T}}^{k+1}$ 
	\end{algorithmic}
\end{algorithm} 

\section{Experimental Results}
\label{E}
In this section, the proposed QT-RNNS1 and QT-RNNS2 methods are compared with seven related methods. Subsection \ref{es} lists the experimental settings. Subsection \ref{er} presents the color video recovery results. Subsection \ref{d} provides a discussion of the experimental results. 
\subsection{Experimental Settings} \label{es}
\subsubsection{Compared Methods} TNN \cite{DBLP:journals/tsp/ZhangA17},  TNN-log \cite{ji2017non}, TNN-TV \cite{DBLP:conf/aaai/LiYX17}, TMac-inc and TMac-dec \cite{xu2015parallel}, TMac-TT\cite{DBLP:journals/tip/BenguaPTD17}, R-TNN\cite{DBLP:conf/icpr/XueQLJ18}, LRQTC \cite{DBLP:journals/pr/MiaoKL20}. The differences and relationships among all methods are listed in Table \ref{t}.
\renewcommand{\arraystretch}{1.3}
\begin{table*}
	\centering
	\fontsize{6.5}{8}\selectfont
	\begin{threeparttable}
		\caption{The differences and relationships of the related methods.}
		\begin{tabular}{p{1.5cm}p{2.5cm}p{6.5cm}p{6cm}}
				\hline
				Method & Representation of video& Description of  low-rankness& Other prior information\\
				\hline
				TNN &tensor-based &tensor nuclear norm based on tensor SVD&-
				\\
				TNN-log& tensor-based &sum of weighted logarithmic norm of mode-n unfolding matrices &-
				\\
				TNN-TV &tensor-based &sum of  nuclear norm based on Tucker decomposition& total variation regularization for local piecewise smooth structure
				\\
				T-TNN &tensor-based &truncated tensor nuclear norm based on tensor SVD&-
				\\
				TMac&tensor-based&low-rank matrix factorization based on Tucker decomposition& -
				\\
				TMac-TT&tensor-based& low-rank matrix factorization based on Tensor Train decomposition& -
				\\
				LRQTC&quaternion tensor-based& sum of  weighted nuclear norm based on Tucker decomposition&-
				\\
				\textbf{ QT-RNNS1}&\textbf{quaternion tensor-based}&  \textbf{truncated nuclear norm based on TQt-SVD}&\textbf{sparse regularization based on QTDCT}
				\\
				\textbf{QT-RNNS2}&\textbf{quaternion tensor-based} & \textbf{truncated nuclear norm with logarithmic norm based on TQt-SVD}&\textbf{sparse regularization based on QTDCT} \\
				\hline
			\end{tabular}\label{t}
	\end{threeparttable}
\end{table*}

\subsubsection{Test Data and Experimental Facility}
Eight color videos were selected, including\textit{ bus, akiyo,  stefan, tempete, waterfall, suzie, container, and grandma} \footnote{http://trace.eas.asu.edu/yuv/}. In the following experiments, only the first 20 frames are adopted, such that the first six videos are of size $288\times352\times3\times20$ and the last three videos are of size $144\times176\times3\times20$. They are reconstructed as pure quaternion tensors with sizes $288\times352\times20$ and $144\times176\times20$, respectively. The Sample Rates (SRs)  are set to  50\%, 30\%, 20\%, 10\%, respectively.
All the experiments were implemented in MATLAB R2019a, on a PC with a 3.00GHz CPU and 8GB RAM.

\subsubsection{Evaluation Indicators and Parameters Setting}
Two indicators are employed to evaluate the performance of the proposed methods, including peak signal to noise rate (PSNR)  and average structural similarity index (ASSIM). The parameters of  QT-RNNS1 were set as $\beta^1=1e-1, \lambda=0.05, \rho=1.1$, and the parameters of QT-RNNS2 were set as $\beta^1=1e-1, \lambda=0.05, \rho=1.01$. Besides, based on the author's codes, the parameters of other compared methods are adjusted to the best.

\subsection{Color Video Recovery Results} \label{er}
The PSNR and ASSIM are listed in Table \ref{ta1} and Table \ref{t2}, and are obtained by all compared methods. The best numerical results are \textbf{bolded}. It can be observed that the proposed QT-RNNS2 can achieve the best results in most cases, except when SR$=0.5$ and SR$=0.3$, the  QT-RNNS1 obtained the best PSNR values when recovering \textit{bus}. Although QT-RNNS1 method is inferior to some tensor-based methods in some cases, the results are acceptable when compared to the quaternion-based methods . 

\begin{figure}[htbp]
	\captionsetup{font={footnotesize}}
	\subfigure{
		\begin{minipage}{18cm}
			\includegraphics[scale=.45]{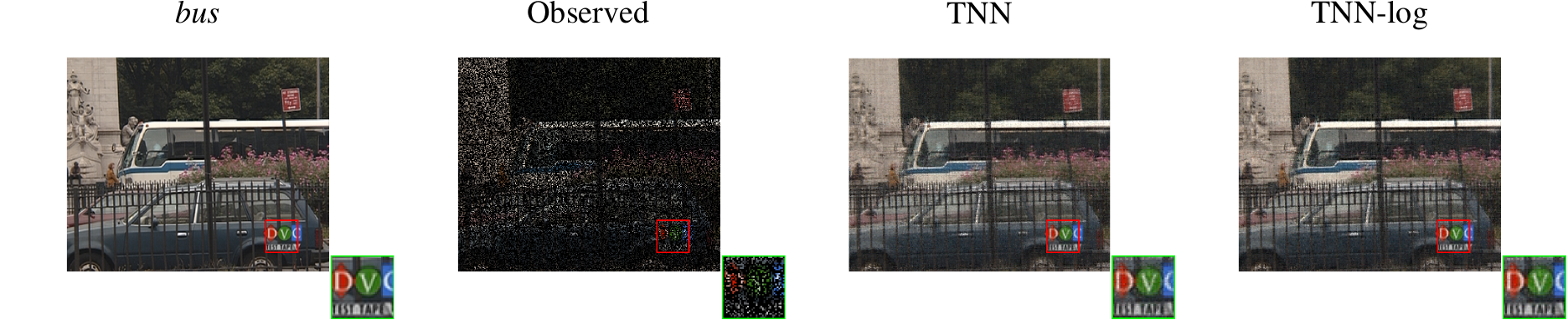}
		\end{minipage}
	}
	\subfigure{
		\begin{minipage}{18cm}
			\includegraphics[scale=.45]{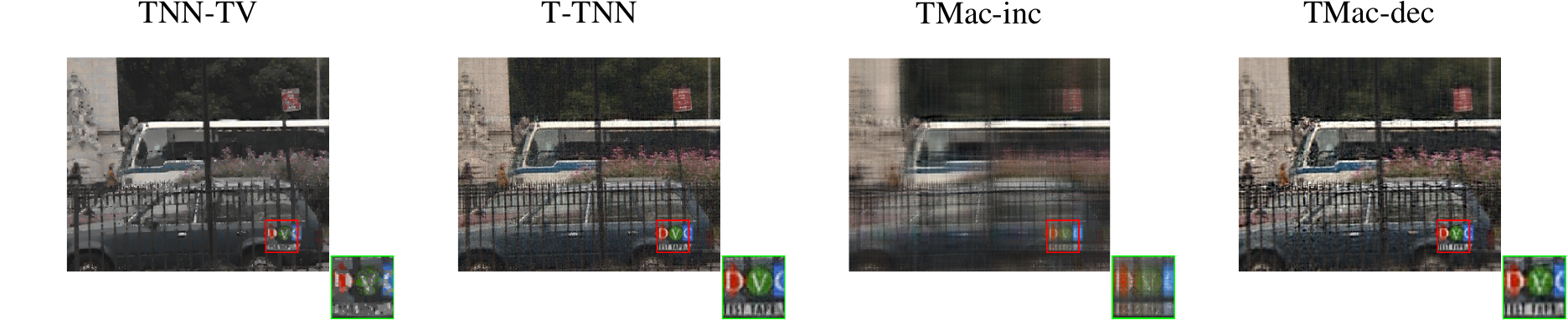}
		\end{minipage}
	}
	\subfigure{
		\begin{minipage}{18cm}
			\includegraphics[scale=.45]{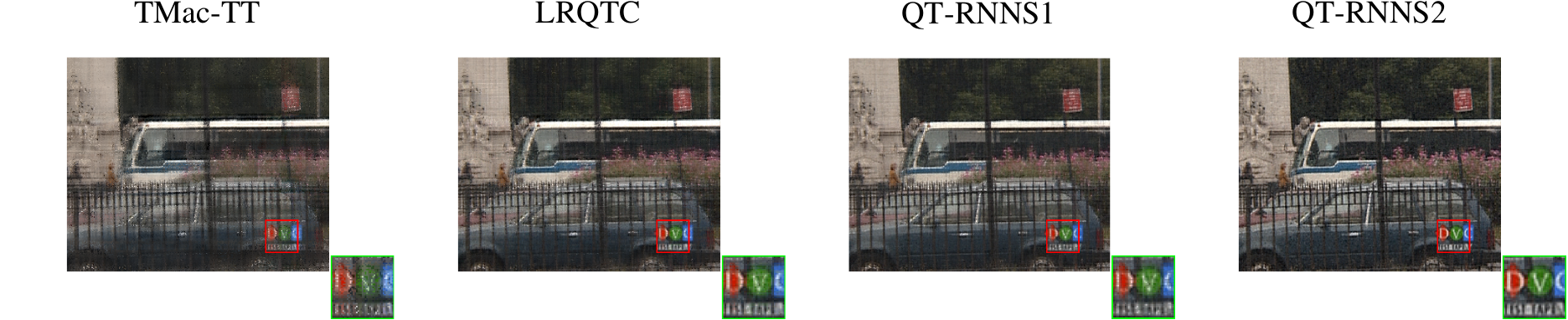}
		\end{minipage}
	}
	\caption{The visual results of  recovering \textit{bus} with SR$=0.3$ at the 18\textit{th} frame.}
	\label{7b}
\end{figure}

\begin{figure}[htbp]
	\captionsetup{font={footnotesize}}
	\subfigure{
		\begin{minipage}{18cm}
			\includegraphics[scale=.45]{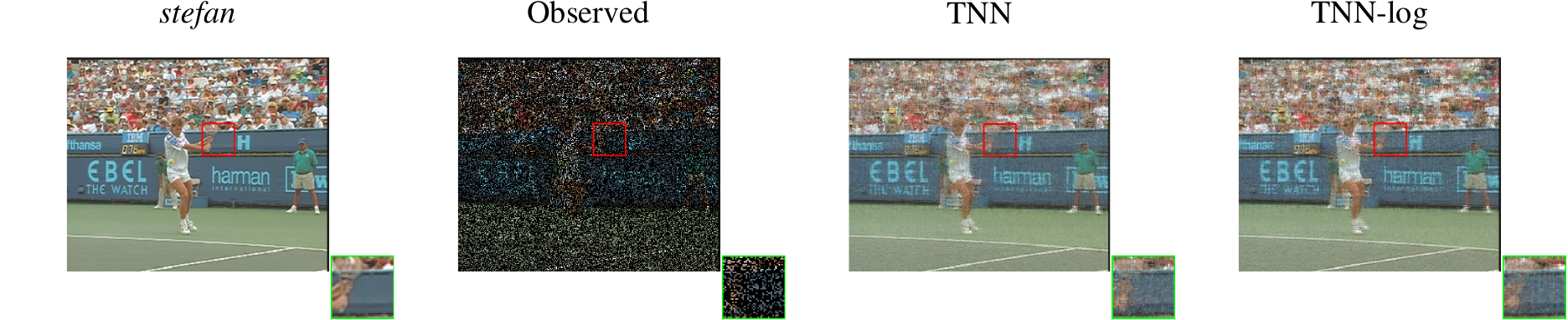}
		\end{minipage}
	}
	\subfigure{
		\begin{minipage}{18cm}
			\includegraphics[scale=.45]{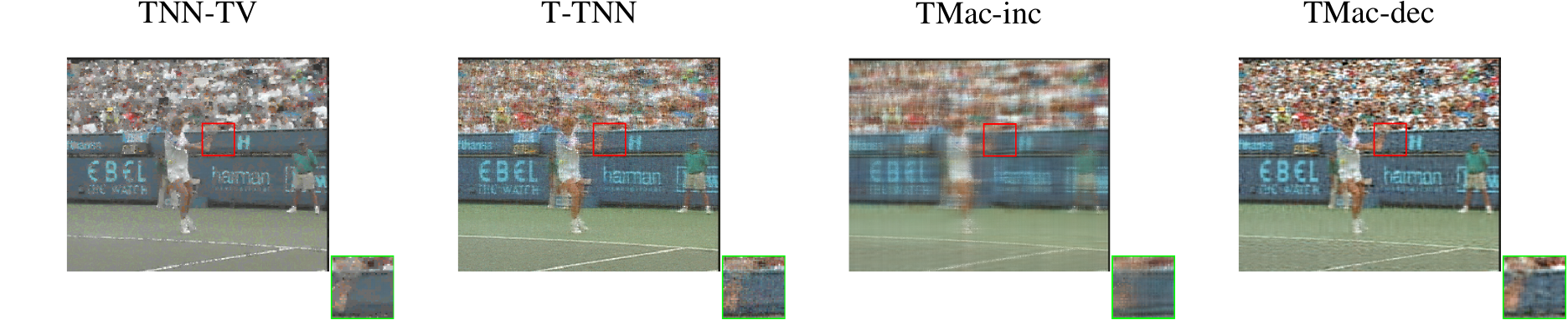}
		\end{minipage}
	}
	\subfigure{
		\begin{minipage}{18cm}
			\includegraphics[scale=.45]{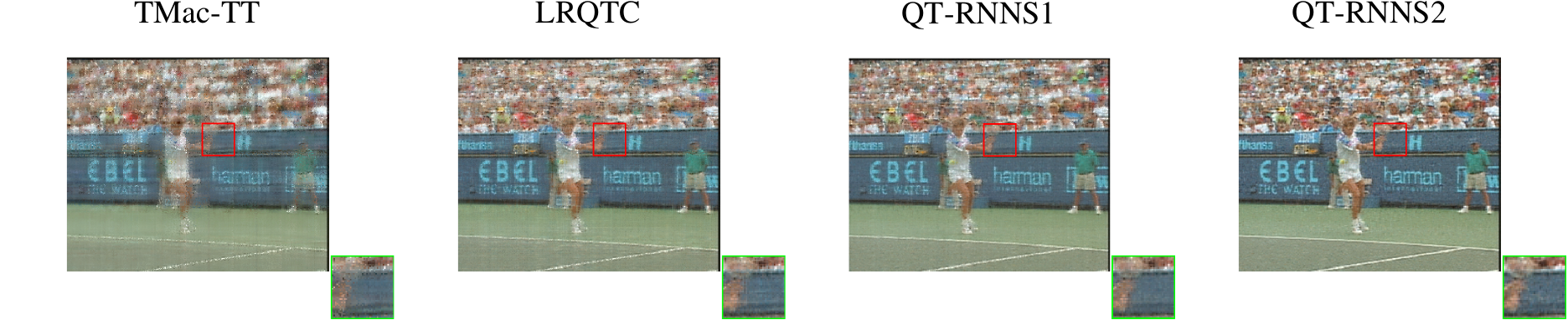}
		\end{minipage}
	}
	\caption{The visual results of  recovering \textit{stefan} with SR$=0.3$ at the 18\textit{th} frame.}
	\label{7s}
\end{figure}

\begin{table*}[htbp] 
	\centering  
	\caption{The $\frac{\text{PSNR}}{\text{ASSIM}}$comparison of different recovery methods for 8 color videos}  
	\label{ta1}
	 \scalebox{.9}{
		\begin{tabular}{|c|c|c|c|c|c|c|c|c|c|c|}
			\hline 
			\multirow{2}{*}{Video}&  
			\multicolumn{10}{c|}{$\text{SR}=0.5$}\cr\cline{2-11}  
			&TNN&TNN-log&TNN-TV& T-TNN&TMac-inc& TMac-dec& TMac-TT& LRQTC& QT-RNNS1&QT-RNNS2\cr \hline  
		$\textit{bus}$ &\makecell{27.4688\\0.8472} 
		&\makecell{27.8489 \\0.8520}&\makecell{26.7656 \\0.8848} &\makecell{27.5426\\0.8481 }&\makecell{22.5015 \\0.7017}  &\makecell{26.0222 \\0.8118}  &\makecell{23.5855 \\0.7589} &\makecell{27.0989\\0.8767}&\makecell{28.1625\\\textbf{0.8916} } &\makecell{\textbf{29.3049}\\0.8792}\cr\hline 
		$\textit{akiyo}$&\makecell{42.0237 \\0.9958} &\makecell{42.2329\\0.9959 }&\makecell{31.7104 \\0.9647 } &\makecell{41.8984 \\0.9953 }&\makecell{29.4959 \\0.9293 } &\makecell{37.3297  \\0.9777 }  &\makecell{32.2096  \\0.9477 }  &\makecell{37.2693  \\0.9846 } &\makecell{40.1668 \\0.9905  }&\pmb{\makecell{47.2225 \\0.9983}} \cr\hline 
		$\textit{stefan}$&\makecell{25.1503\\0.9004 } 
		&\makecell{25.4682 \\0.9052}&\makecell{25.4945\\0.9068} &\makecell{25.1899\\0.9009}
	&\makecell{21.5793  \\0.8030}  &\makecell{25.6508 \\0.9103}  &\makecell{22.4252 \\0.8356} &\makecell{26.4100 \\0.9294 }&\makecell{27.5370 \\0.9454 } &\pmb{\makecell{28.9896\\0.9522}}\cr\hline 
		$\textit{tempete}$&\makecell{29.9098 \\0.9349} &\makecell{30.2200 \\0.9373}&\makecell{25.6139  \\0.8883 } &\makecell{29.9513 \\0.9349  }&\makecell{23.2665  \\0.7831 }  &\makecell{26.8717  \\0.8783 }  &\makecell{24.5690 \\0.8352 } &\makecell{27.8278 \\0.9162 }&\makecell{28.9693 \\0.9311 } &\pmb{\makecell{32.3476 \\0.9583}}\cr\hline   
		$\textit{waterfall}$&\makecell{36.2144 \\0.9837} &\makecell{36.4523  \\0.9844}&\makecell{29.4415 \\0.9358 } &\makecell{36.2379 \\0.9837 }&\makecell{28.0918 \\0.9077 }  &\makecell{31.7318  \\0.9543 }  &\makecell{29.4339 \\0.9318 } &\makecell{32.6245\\ 0.9678   }&\makecell{35.1978\\0.9802  } &\pmb{\makecell{39.8446 \\0.9921}}\cr\hline 
		$\textit{suzie}$&\makecell{37.0243\\0.9844} &\makecell{37.2051\\0.9849}&\makecell{32.1941 \\0.9515 } &\makecell{36.9587\\0.9840}&\makecell{31.9933\\0.9531}  &\makecell{20.8712  \\0.8327}  &\makecell{34.1747 \\0.9683 } &\makecell{34.6803 \\0.9756  }&\makecell{35.4444\\0.9773 } &\pmb{\makecell{40.1347 \\0.9908}}\cr\hline 
		$\textit{container}$&\makecell{39.2179\\0.9924} &\makecell{39.4946\\0.9927}&\makecell{26.5751 \\0.9202 } &\makecell{38.8865 \\0.9918}&\makecell{27.9132  \\0.9249 }  &\makecell{18.4649  \\0.7712 }  &\makecell{31.6999 \\0.9635 } &\makecell{30.9900 \\0.9668  }&\makecell{32.4229 \\0.9722  } &\pmb{\makecell{47.3978 \\0.9974}}\cr\hline  
		$\textit{grandma}$&\makecell{46.2940\\0.9986  } &\makecell{46.4175 \\0.9986}&\makecell{30.0921  \\0.9540 } &\makecell{46.1753 \\0.9986}&\makecell{31.3900  \\0.9613  }  &\makecell{24.5059  \\0.9213  }  &\makecell{36.2686  \\0.9862 } &\makecell{36.1837  \\0.9877 }&\makecell{36.6203 \\0.9889 } &\pmb{\makecell{47.9809 \\0.9990}}\cr\hline 
			\hline  
			\multirow{1}{*}{Video}&  
			\multicolumn{10}{c|}{$\text{SR}=0.3$}\cr\cline{2-11}  
			&TNN&TNN-log&TNN-TV& T-TNN&TMac-inc& TMac-dec& TMac-TT& LRQTC& QT-RNNS1&QT-RNNS2\cr \hline  
			$\textit{bus}$ &\makecell{23.7394\\0.6998} 
			&\makecell{24.2432\\0.7123 }&\makecell{23.2642 \\0.7453} &\makecell{23.8325\\0.7016 }&\makecell{22.9045 \\0.6080}  &\makecell{21.7801 \\0.6850}  &\makecell{23.4549 \\0.6487} &\makecell{24.5961\\0.7331}&\makecell{25.7786\\\textbf{0.7823}} &\makecell{\textbf{29.3049}\\0.7713}\cr\hline 
			$\textit{akiyo}$&\makecell{38.1462 \\0.9900 } &\makecell{38.5320 \\0.9907 }&\makecell{22.7441  \\0.8489 } &\makecell{37.8275\\0.9889 }&\makecell{28.3054  \\0.9091 } &\makecell{32.7141  \\0.9503  }  &\makecell{30.0190   \\0.9192 }  &\makecell{32.2587  \\0.9562  } &\makecell{34.2534  \\0.9705  }&\pmb{\makecell{42.9540  \\0.9960}} \cr\hline 
			$\textit{stefan}$&\makecell{21.7232\\0.7888} 
			&\makecell{21.9902 \\0.7966 }&\makecell{21.5849\\0.7189} &\makecell{21.7321\\0.7888}
			&\makecell{20.2946 \\0.7350}  &\makecell{22.4055\\0.8318}  &\makecell{20.6749\\0.7541} &\makecell{22.3532 \\0.8264  }&\makecell{23.4745 \\0.8662} &\pmb{\makecell{24.6364 \\0.8807}}\cr\hline 
			$\textit{tempete}$&\makecell{26.1205 \\0.8581} &\makecell{26.5584\\0.8657 }&\makecell{21.4621\\0.7217} &\makecell{26.1117 \\0.8552}&\makecell{21.9732 \\0.7152}  &\makecell{24.3118  \\0.7991}  &\makecell{22.7670 \\0.7568 } &\makecell{24.4188 \\0.8156  }&\makecell{25.2774  \\0.8536} &\pmb{\makecell{28.6686  \\0.9113}}\cr\hline   
			$\textit{waterfall}$&\makecell{32.3438\\0.9620} &\makecell{32.7389 \\0.9647 }&\makecell{22.1695\\0.7371} &\makecell{32.2264 \\0.9607}&\makecell{26.8074 \\0.8792 }  &\makecell{26.8360  \\0.9064 }  &\makecell{27.6290 \\0.8973} &\makecell{29.0284 \\ 0.9288   }&\makecell{30.8075 \\0.9483} &\pmb{\makecell{35.9844 \\0.9816 }}\cr\hline 
			$\textit{suzie}$&\makecell{33.3655\\0.9668 } &\makecell{33.7148 \\0.9686 }&\makecell{26.9049 \\0.7988 } &\makecell{32.9049 \\0.9631 }&\makecell{30.4892 \\0.9366 }  &\makecell{15.8579 \\0.5306 }  &\makecell{31.9480  \\0.9492 } &\makecell{30.8670 \\0.9498   }&\makecell{31.3812 \\0.9496  } &\pmb{\makecell{36.6677  \\0.9809 }}\cr\hline 
			$\textit{container}$&\makecell{35.2095\\0.9823 } &\makecell{35.7235\\0.9837 }&\makecell{23.0744  \\0.7725 } &\makecell{34.4224\\0.9800}&\makecell{26.5851  \\0.9001   }  &\makecell{12.4708  \\0.5230  }  &\makecell{29.0497 \\0.9360  } &\makecell{26.5512 \\0.9151  }&\makecell{27.1275 \\0.9217   } &\pmb{\makecell{42.3936  \\0.9939 }}\cr\hline  
			$\textit{grandma}$&\makecell{43.0166\\0.9971} &\makecell{43.3858\\0.9973 }&\makecell{22.8230  \\0.7525 } &\makecell{42.6545\\0.9968  }&\makecell{30.1309  \\0.9488  }  &\makecell{19.4189   \\0.7586  }  &\makecell{33.5095  \\0.9742 } &\makecell{31.4801  \\0.9669    }&\makecell{31.5416  \\0.9683 } &\pmb{\makecell{45.1272  \\0.9980 }}\cr\hline 
	\end{tabular}}
\end{table*}  

\begin{table*}[htbp] 
	\centering  
	\caption{The $\frac{\text{PSNR}}{\text{ASSIM}}$ comparison of different recovery methods for 8 color videos}  
	\label{t2}
	\scalebox{.9}{
		\begin{tabular}{|c|c|c|c|c|c|c|c|c|c|c|}
			\hline 
			\multirow{2}{*}{Video}&  
			\multicolumn{10}{c|}{$\text{SR}=0.2$}\cr\cline{2-11}  
			&TNN&TNN-log&TNN-TV& T-TNN&TMac-inc& TMac-dec& TMac-TT& LRQTC& QT-RNNS1&QT-RNNS2\cr \hline  
			$\textit{bus}$ &\makecell{21.8280\\0.5931} 
			&\makecell{22.3784 \\0.6113}&\makecell{21.5413 \\0.6319} &\makecell{21.9307\\0.5953 }&\makecell{20.4624\\0.5495}  &\makecell{18.4727 \\0.4981}  &\makecell{20.9389\\0.5807} &\makecell{21.6094\\0.6346}&\makecell{22.7354\\0.6967} &\pmb{\makecell{24.0853\\0.6986}}\cr\hline 
			$\textit{akiyo}$&\makecell{35.9674 \\0.9839 } &\makecell{36.5191\\0.9854 }&\makecell{19.1079  \\0.6944 } &\makecell{35.5094\\0.9815 }&\makecell{27.7474  \\0.8968  } &\makecell{22.2224  \\0.7897 }  &\makecell{29.0119  \\0.9023 }  &\makecell{29.5693  \\0.9285 } &\makecell{31.2501  \\0.9481  }&\pmb{\makecell{40.4373 \\0.9983}} \cr\hline 
			$\textit{stefan}$&\makecell{21.7232\\0.7888 } 
			&\makecell{21.9902 \\0.7966}
			&\makecell{19.7294 \\0.5349 } &\makecell{21.7321 \\0.7888}
			&\makecell{19.6133   \\0.6902 }  &\makecell{17.4463  \\0.6549 }  &\makecell{19.8573  \\0.7034 } &\makecell{20.4872  \\0.7448  }&\makecell{21.4389  \\0.7934  } &\pmb{\makecell{22.5682 \\0.8192}}\cr\hline 
			$\textit{tempete}$&\makecell{26.1205  \\0.8581} &\makecell{26.5584 \\0.8657 }&\makecell{19.4599  \\0.5773 } &\makecell{26.1117 \\0.8552 }&\makecell{21.3331   \\0.6756  }  &\makecell{20.8422  \\0.6654 }  &\makecell{21.9784  \\0.7121  } &\makecell{22.7217  \\0.7473  }&\makecell{23.3364  \\0.7855  } &\pmb{\makecell{26.6932 \\0.8700 }}\cr\hline   
			$\textit{waterfall}$&\makecell{32.3438  \\0.9620 } &\makecell{32.7389 \\0.9647}&\makecell{19.1428  \\0.5020  } &\makecell{17.7733  \\0.3313   }&\makecell{32.2264\\0.9607 }  &\makecell{22.8920   \\0.8426 }  &\makecell{26.8652  \\0.8781 } &\makecell{27.0465 \\ 0.8924  }&\makecell{28.6831 \\0.9184 } &\pmb{\makecell{33.6559 \\0.9697}}\cr\hline 
			$\textit{suzie}$&\makecell{31.4173 \\0.9507} &\makecell{31.9273  \\0.9547 }&\makecell{24.4801  \\0.6262 } &\makecell{30.6578 \\0.9415}&\makecell{29.6985 \\0.9253 }  &\makecell{12.6942  \\0.2244 }  &\makecell{30.7311  \\0.9345 } &\makecell{28.6060  \\0.9145  }&\makecell{28.9900 \\0.9222 } &\pmb{\makecell{34.6850  \\0.9706 }}\cr\hline 
			$\textit{container}$&\makecell{32.7163\\0.9710} &\makecell{33.4670  \\0.9744  }&\makecell{21.3214  \\0.6191 } &\makecell{31.5467\\0.9639 }&\makecell{25.7684  \\0.8817  }  &\makecell{9.9466  \\0.1523 }  &\makecell{27.5190  \\0.9130  } &\makecell{24.1345  \\0.8566  }&\makecell{24.5348  \\0.8703  } &\pmb{\makecell{38.6235 \\0.9885 }}\cr\hline  
			$\textit{grandma}$&\makecell{19.0741  \\0.2940  } &\makecell{19.0728  \\0.2956  }&\makecell{19.7285  \\0.4841 } &\makecell{19.0706  \\0.2930} &\makecell{29.4595  \\0.9400 }  &\makecell{16.5534   \\0.5292 }  &\makecell{32.0661  \\0.9646  } &\makecell{28.6993  \\0.9414  }&\makecell{28.8418   \\0.9462  } &\pmb{\makecell{43.2002  \\0.9970 }}\cr\hline 
			\hline  
			\multirow{2}{*}{Video}&  
			\multicolumn{10}{c|}{$\text{SR}=0.1$}\cr\cline{2-11}  
			&TNN&TNN-log&TNN-TV& T-TNN&TMac-inc& TMac-dec& TMac-TT& LRQTC& QT-RNNS1&QT-RNNS2\cr \hline  
			$\textit{bus}$ &\makecell{19.6664\\0.4522} 
			&\makecell{19.9377\\0.4590}
			&\makecell{19.5717 \\0.4828} &\makecell{19.4149\\0.4350}
			&\makecell{19.5916\\0.4686}  &\makecell{14.0292 \\0.1959}  &\makecell{19.7682 \\0.4854} &\makecell{19.3573\\0.4897}&\makecell{19.4710\\0.5314} &\pmb{\makecell{21.9361\\0.5863}}\cr\hline 
			$\textit{akiyo}$&\makecell{32.9523\\0.9666} &\makecell{32.9349 \\0.9647}&\makecell{17.7276   \\0.5682 } &\makecell{31.8867 \\0.9556 }&\makecell{27.0762  \\0.8812  } &\makecell{13.8732 \\0.4587  }  &\makecell{27.9046   \\0.8801 }  &\makecell{25.9348 \\0.8651  } &\makecell{24.9418   \\0.8469  }&\pmb{\makecell{36.5933  \\0.9845}} \cr\hline 
			$\textit{stefan}$&\makecell{18.5912 \\0.6022 } 
			&\makecell{18.5178\\0.5964 }&\makecell{18.5187 \\0.5606 } &\makecell{18.1608 \\0.5780}
			&\makecell{18.7304 \\0.6288 }  &\makecell{12.6956 \\0.2878 }  &\makecell{18.7689 \\0.6317 } &\makecell{18.4741 \\0.6181  }&\makecell{17.8973  \\0.6242 } &\pmb{\makecell{18.5270  \\0.6359}}\cr\hline 
			$\textit{tempete}$&\makecell{21.9430\\0.6927 } &\makecell{22.1501 \\0.7001}&\makecell{17.6936 \\0.4100 } &\makecell{21.2944 \\0.6485 }&\makecell{20.6303 \\0.6237 }  &\makecell{14.7226  \\0.3087 }  &\makecell{21.1393  \\0.6540  } &\makecell{20.4420 \\0.6324  }&\makecell{19.9407  \\0.6456 } &\pmb{\makecell{24.2331  \\0.7951 }}\cr\hline   
			$\textit{waterfall}$&\makecell{27.5731 \\0.8948} &\makecell{27.6326 \\0.8963 }&\makecell{17.9488 \\0.3211 } &\makecell{26.5949 \\0.8757 }&\makecell{25.4843  \\0.8409 }  &\makecell{16.8750  \\0.5648  }  &\makecell{26.0418  \\0.8546 } &\makecell{24.8760  \\ 0.8355   }&\makecell{25.7330  \\0.8565 } &\pmb{\makecell{30.4565  \\0.9414} }\cr\hline 
			$\textit{suzie}$&\makecell{28.7691 \\0.9193  } &\makecell{29.1793\\0.9234 }&\makecell{22.4736 \\0.4667  } &\makecell{26.5000 \\0.8780 }&\makecell{28.3883  \\0.9057  }  &\makecell{9.9371 \\0.0592 }  &\makecell{28.8264  \\0.9094 } &\makecell{25.0000  \\0.8500  }&\makecell{23.2253  \\0.8156 } &\pmb{\makecell{31.9447   \\0.9496 }}\cr\hline 
			$\textit{container}$&\makecell{28.2328 \\0.9331 } &\makecell{28.1848 \\0.9300 }&\makecell{19.4086   \\0.4668  } &\makecell{25.0618\\0.8808 }&\makecell{23.4926   \\0.8337    }  &\makecell{7.1822   \\0.0333  }  &\makecell{24.8569  \\0.8624  } &\makecell{21.4210 \\0.7439  }&\makecell{18.7180  \\0.6174    } &\pmb{\makecell{31.9465  \\0.9647  }}\cr\hline  
			$\textit{grandma}$&\makecell{15.0825 \\0. 5019} &\makecell{15.0834  \\0.5024  }&\makecell{18.3090 \\0.2774 } &\makecell{29.9678 \\0.9578 }&\makecell{28.2759  \\0.9256  }  &\makecell{13.7822 \\0.2109  }  &\makecell{30.1074  \\0.9476  } &\makecell{24.9565   \\0.8889  }&\makecell{24.6477  \\0.8876 } &\pmb{\makecell{38.5849  \\0.9915}}\cr\hline 
	\end{tabular}}
\end{table*}  

\begin{figure}[htbp]
	\captionsetup{font={footnotesize}}
	\subfigure{
		\begin{minipage}{18cm}
			\includegraphics[scale=.55]{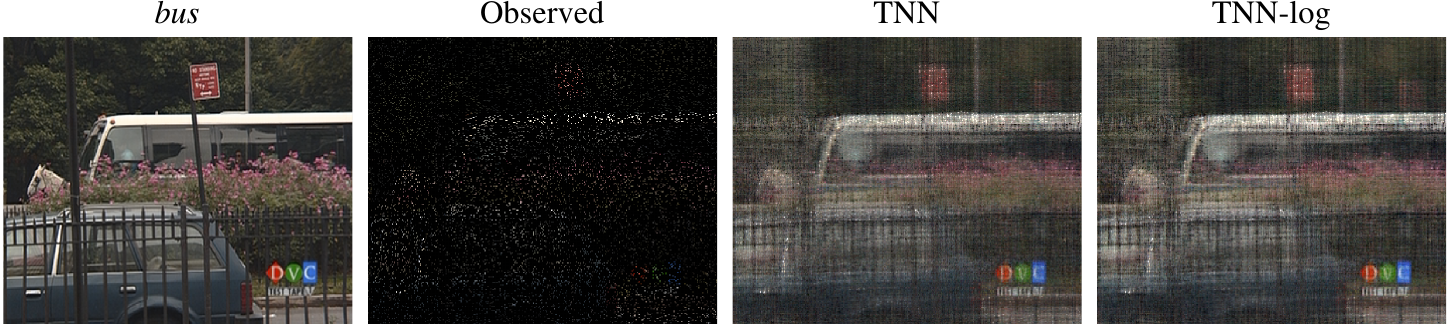}
		\end{minipage}
	}
	\subfigure{
		\begin{minipage}{18cm}
			\includegraphics[scale=.55]{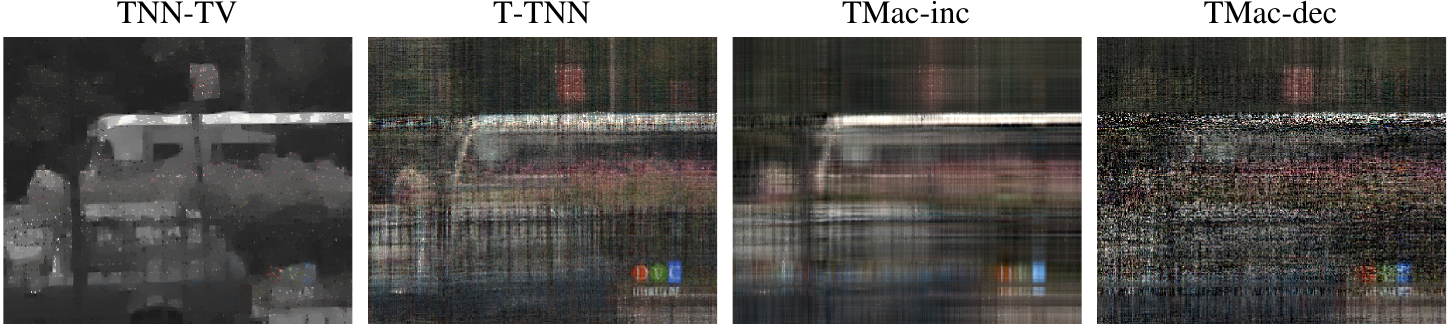}
		\end{minipage}
	}
	\subfigure{
		\begin{minipage}{18cm}
			\includegraphics[scale=.55]{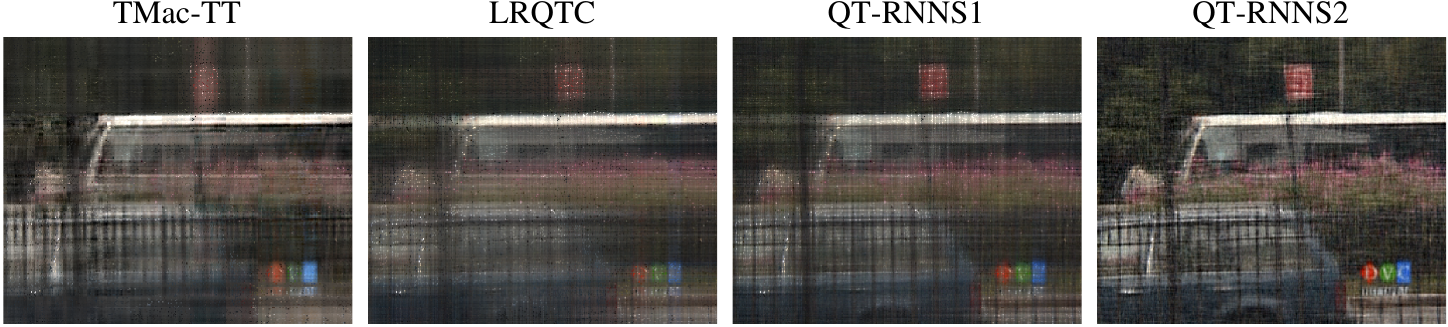}
		\end{minipage}
	}
	\caption{The visual results of  recovering \textit{bus} with SR=0.1 at the first frame.}
	\label{b1}
\end{figure}

\begin{figure}[htbp]
	\captionsetup{font={footnotesize}}
	\subfigure{
		\begin{minipage}{18cm}
			\includegraphics[scale=.55]{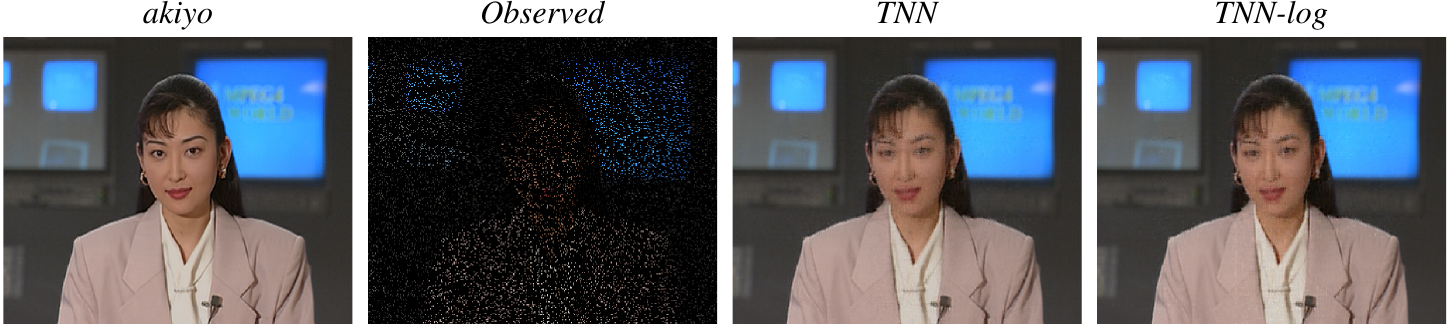}
		\end{minipage}
	}
	\subfigure{
		\begin{minipage}{18cm}
			\includegraphics[scale=.55]{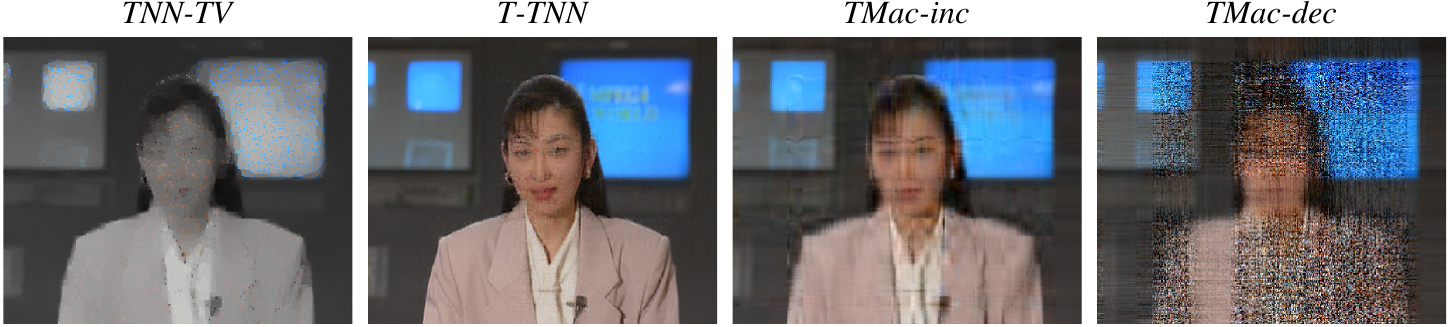}
		\end{minipage}
	}
	\subfigure{
		\begin{minipage}{18cm}
			\includegraphics[scale=.55]{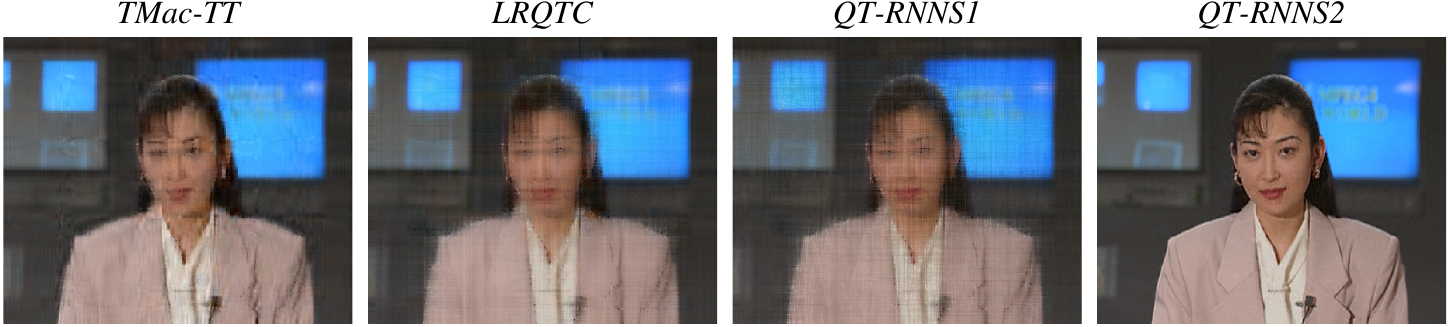}
		\end{minipage}
	}
	\caption{The visual results of  recovering \textit{akiyo} with SR=0.1 at the first frame.}
	\label{ak}
\end{figure}

\begin{figure}[htbp]
	\captionsetup{font={footnotesize}}
	\subfigure{
		\begin{minipage}{18cm}
			\includegraphics[scale=.55]{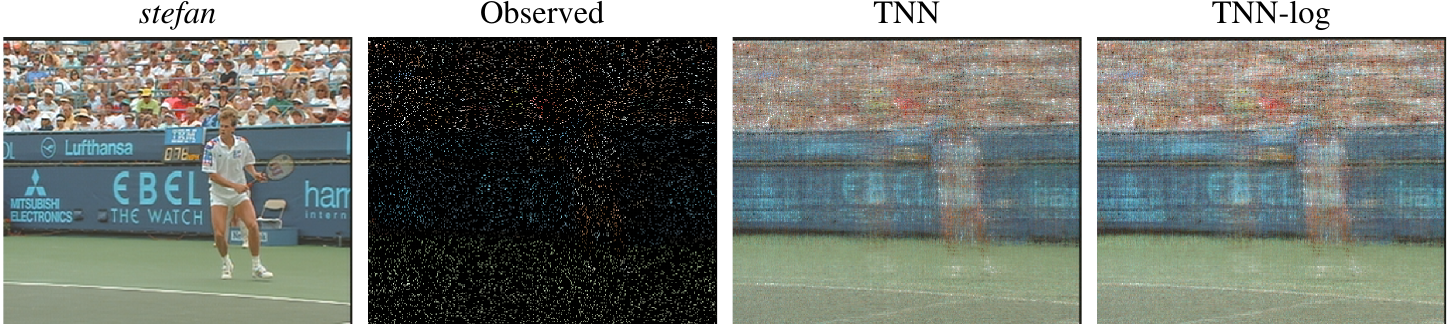}
		\end{minipage}
	}
	\subfigure{
		\begin{minipage}{18cm}
			\includegraphics[scale=.55]{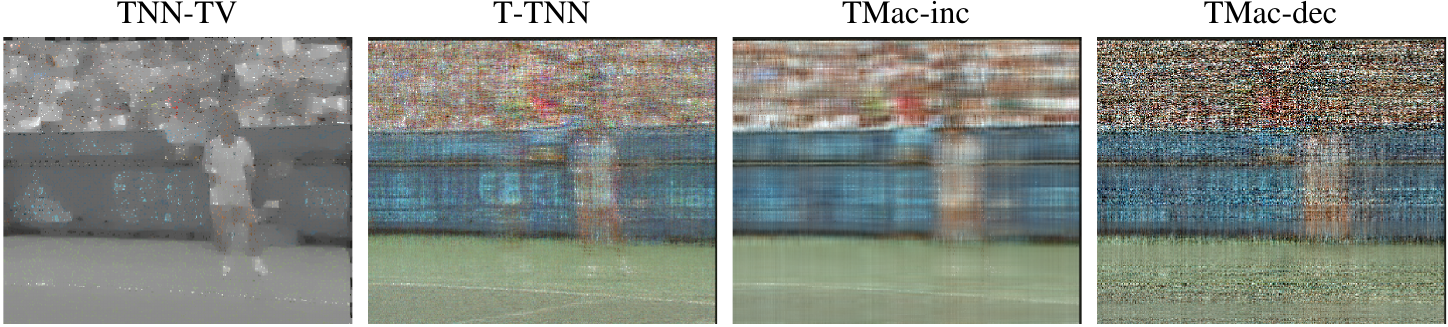}
		\end{minipage}
	}
	\subfigure{
		\begin{minipage}{18cm}
			\includegraphics[scale=.55]{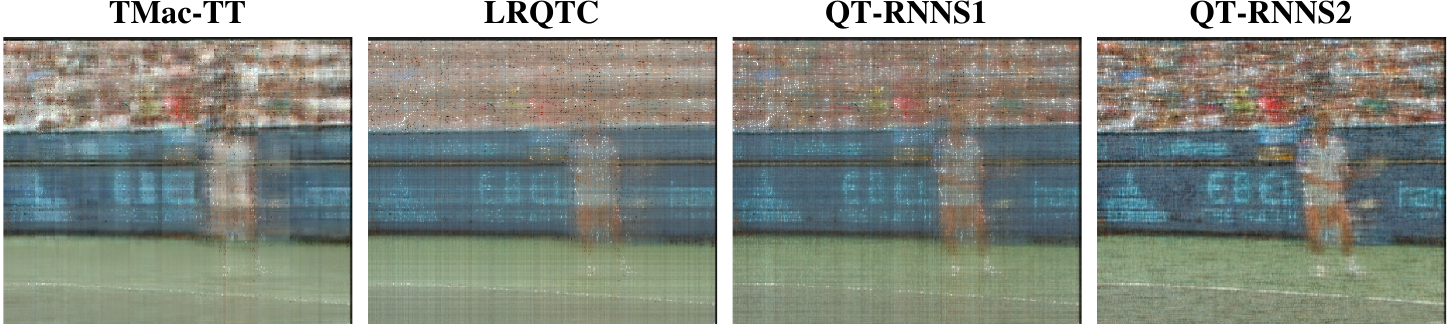}
		\end{minipage}
	}
	\caption{The visual results of  recovering \textit{stefan} with SR=0.1 at the first frame.}
	\label{st}
\end{figure}

\begin{figure}[htbp]
	\captionsetup{font={footnotesize}}
	\subfigure{
		\begin{minipage}{18cm}
			\includegraphics[scale=.55]{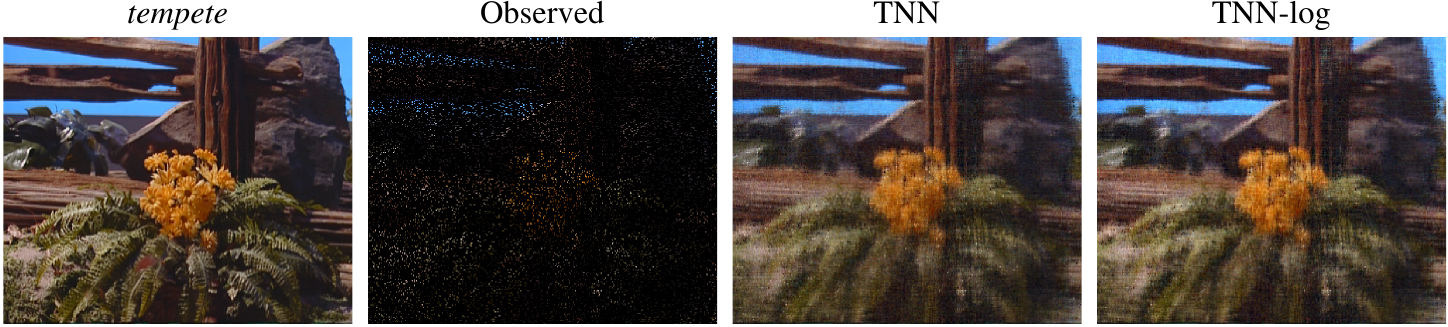}
		\end{minipage}
	}
	\subfigure{
		\begin{minipage}{18cm}
			\includegraphics[scale=.55]{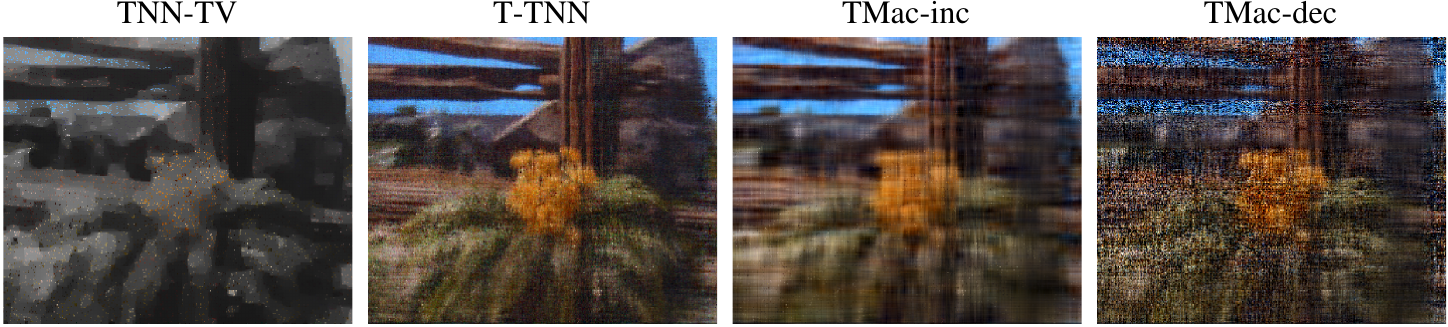}
		\end{minipage}
	}
	\subfigure{
		\begin{minipage}{18cm}
			\includegraphics[scale=.55]{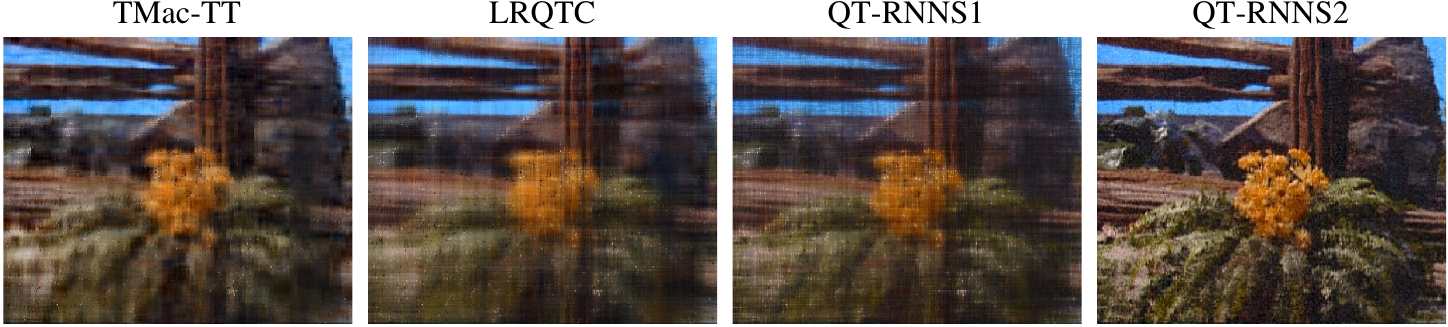}
		\end{minipage}
	}
	\caption{The visual results of  recovering \textit{tempete} with SR=0.1 at the first frame.}
	\label{te}
\end{figure}

\begin{figure}[htbp]
	\captionsetup{font={footnotesize}}
	\subfigure{
		\begin{minipage}{18cm}
			\includegraphics[scale=.55]{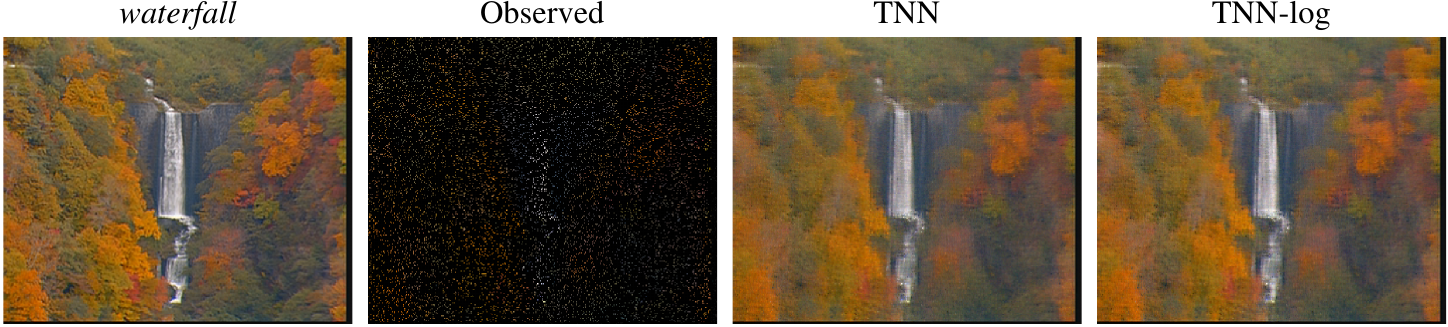}
		\end{minipage}
	}
	\subfigure{
		\begin{minipage}{18cm}
			\includegraphics[scale=.55]{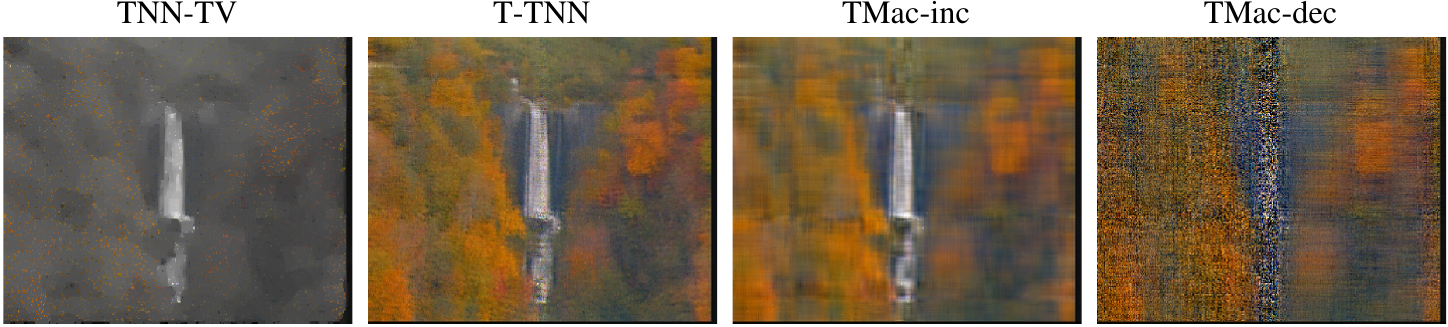}
		\end{minipage}
	}
	\subfigure{
		\begin{minipage}{18cm}
			\includegraphics[scale=.55]{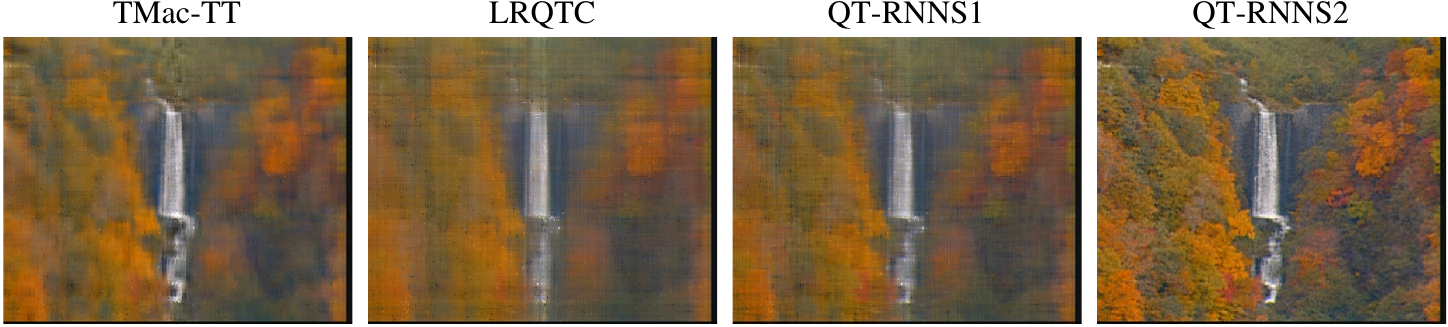}
		\end{minipage}
	}
	\caption{The visual results of  recovering \textit{waterfall} with SR=0.1 at the first frame.}
	\label{wa}
\end{figure}

\begin{figure}[htbp]
	\captionsetup{font={footnotesize}}
	\subfigure{
		\begin{minipage}{18cm}
			\includegraphics[scale=.55]{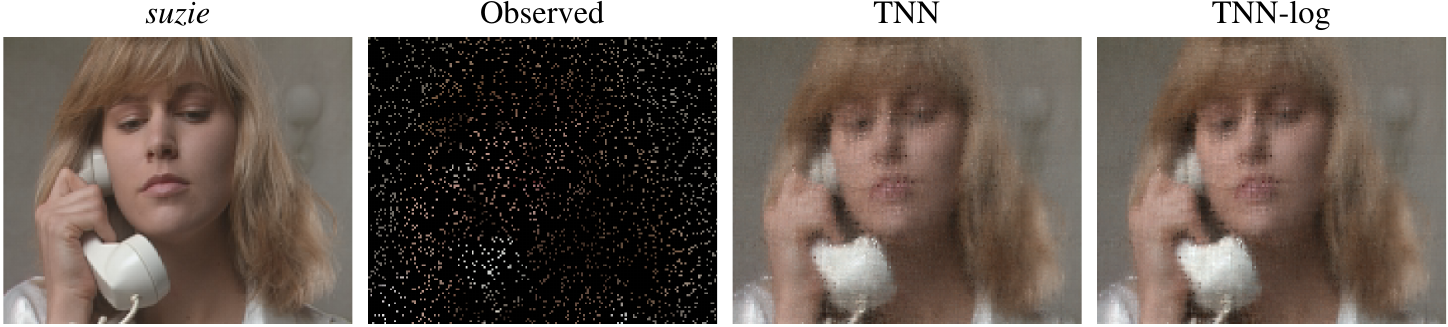}
		\end{minipage}
	}
	\subfigure{
		\begin{minipage}{18cm}
			\includegraphics[scale=.55]{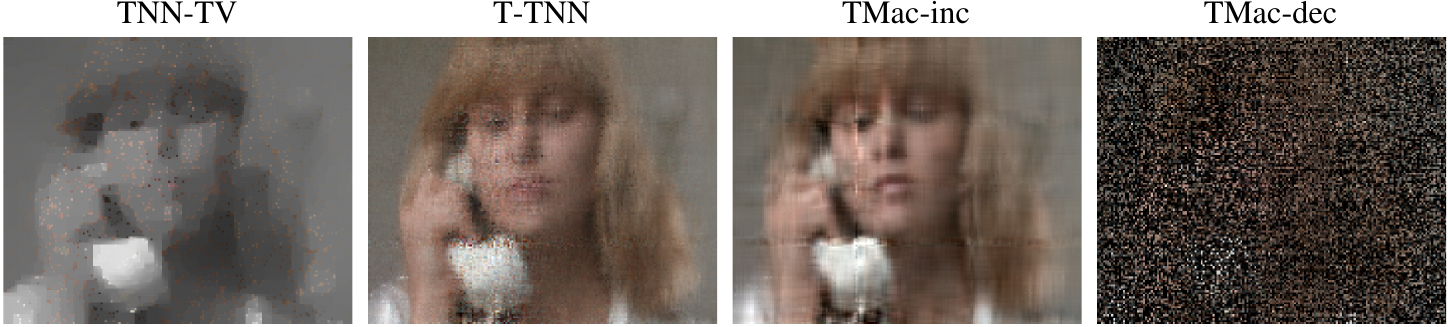}
		\end{minipage}
	}
	\subfigure{
		\begin{minipage}{18cm}
			\includegraphics[scale=.55]{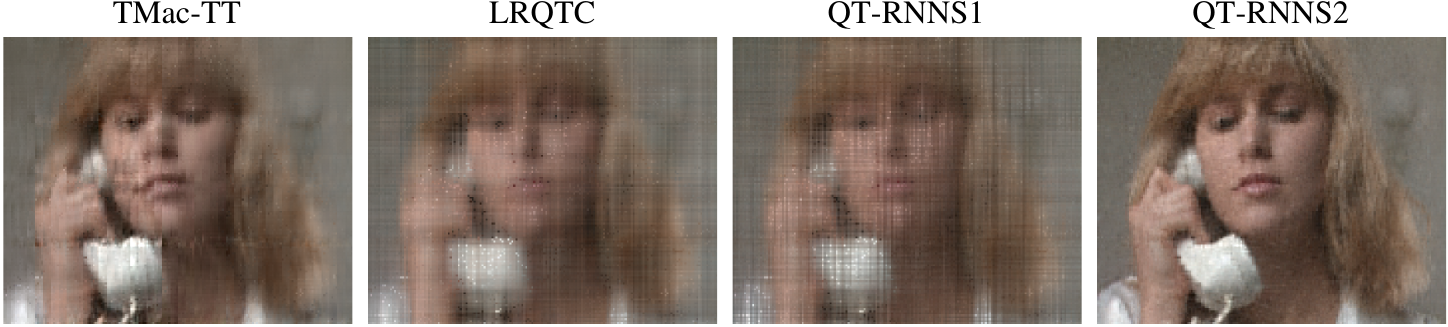}
		\end{minipage}
	}
	\caption{The visual results of  recovering \textit{suzie} with SR=0.1 at the first frame.}
	\label{su}
\end{figure}

\begin{figure}[htbp]
	\captionsetup{font={footnotesize}}
	\subfigure{
		\begin{minipage}{18cm}
			\includegraphics[scale=.55]{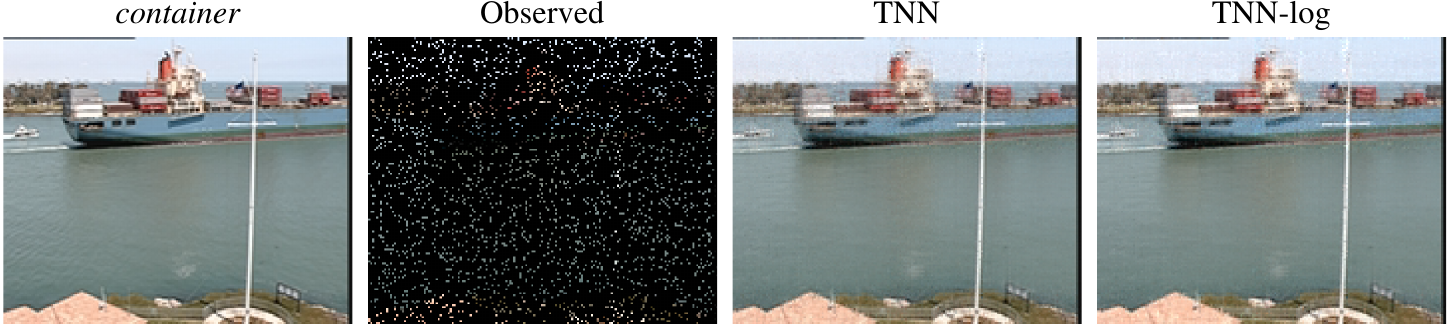}
		\end{minipage}
	}
	\subfigure{
		\begin{minipage}{18cm}
			\includegraphics[scale=.55]{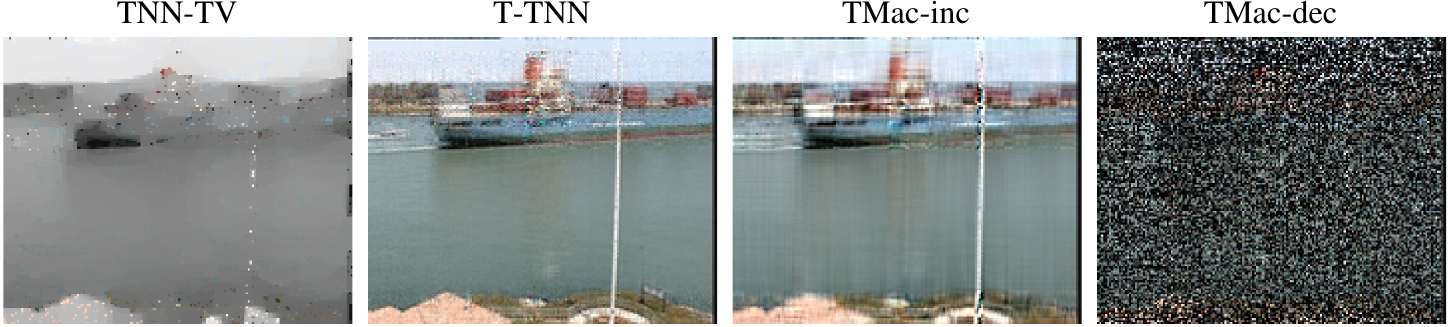}
		\end{minipage}
	}
	\subfigure{
		\begin{minipage}{18cm}
			\includegraphics[scale=.55]{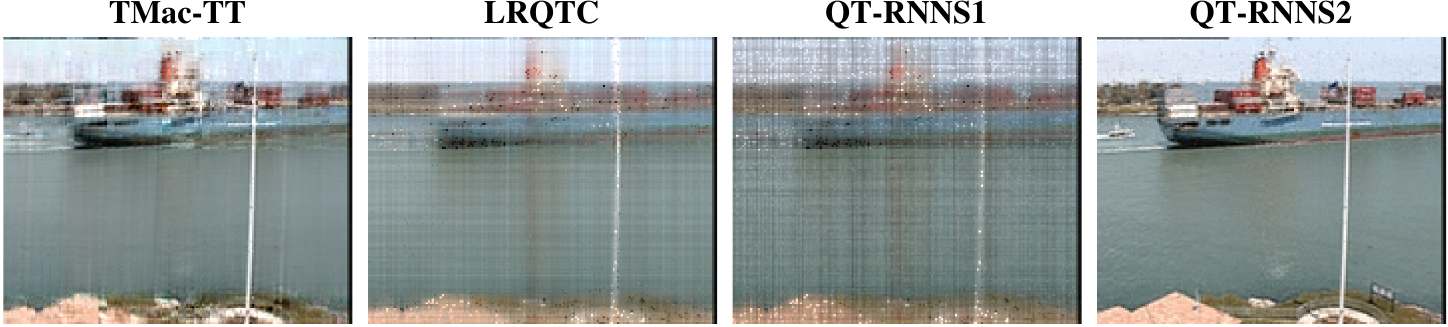}
		\end{minipage}
	}
	\caption{The visual results of  recovering \textit{container} with SR=0.1 at the first frame.}
	\label{c}
\end{figure}

\begin{figure}[htbp]
	\captionsetup{font={footnotesize}}
	\subfigure{
		\begin{minipage}{18cm}
			\includegraphics[scale=.55]{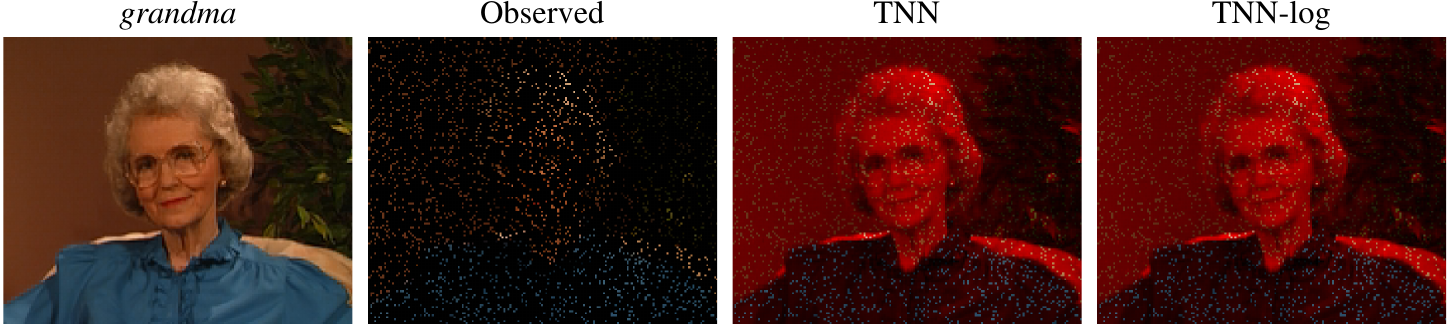}
		\end{minipage}
	}
	\subfigure{
		\begin{minipage}{18cm}
			\includegraphics[scale=.55]{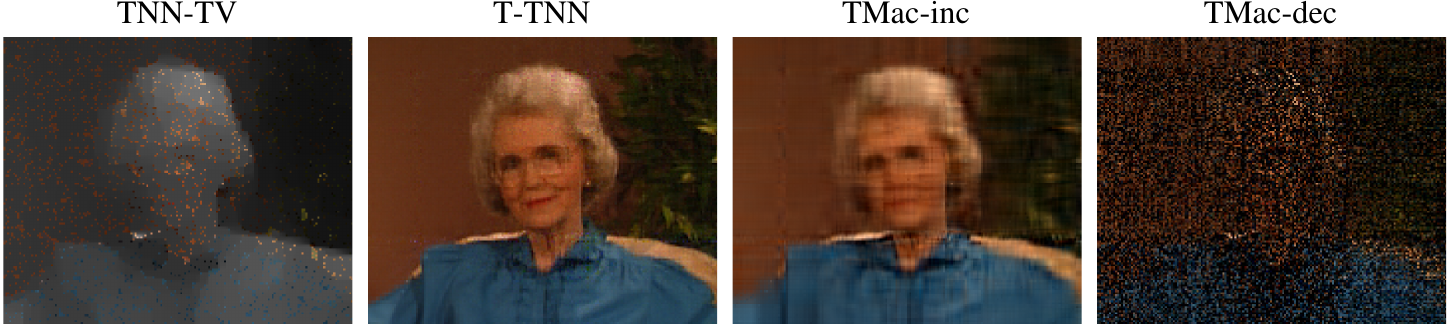}
		\end{minipage}
	}
	\subfigure{
		\begin{minipage}{18cm}
			\includegraphics[scale=.55]{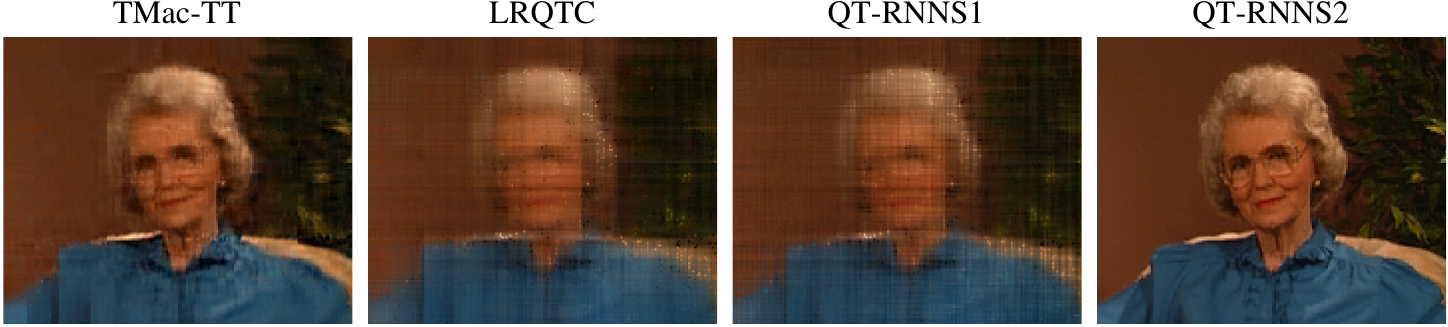}
		\end{minipage}
	}
	\caption{The visual results of  recovering \textit{grandma} with SR=0.1 at the first frame.}
	\label{g}
\end{figure}

\begin{figure}[htbp]
	\captionsetup{font={footnotesize}}
	\subfigure{
		\begin{minipage}{18cm}
			\includegraphics[height = 4.8cm, width = 8.5cm]{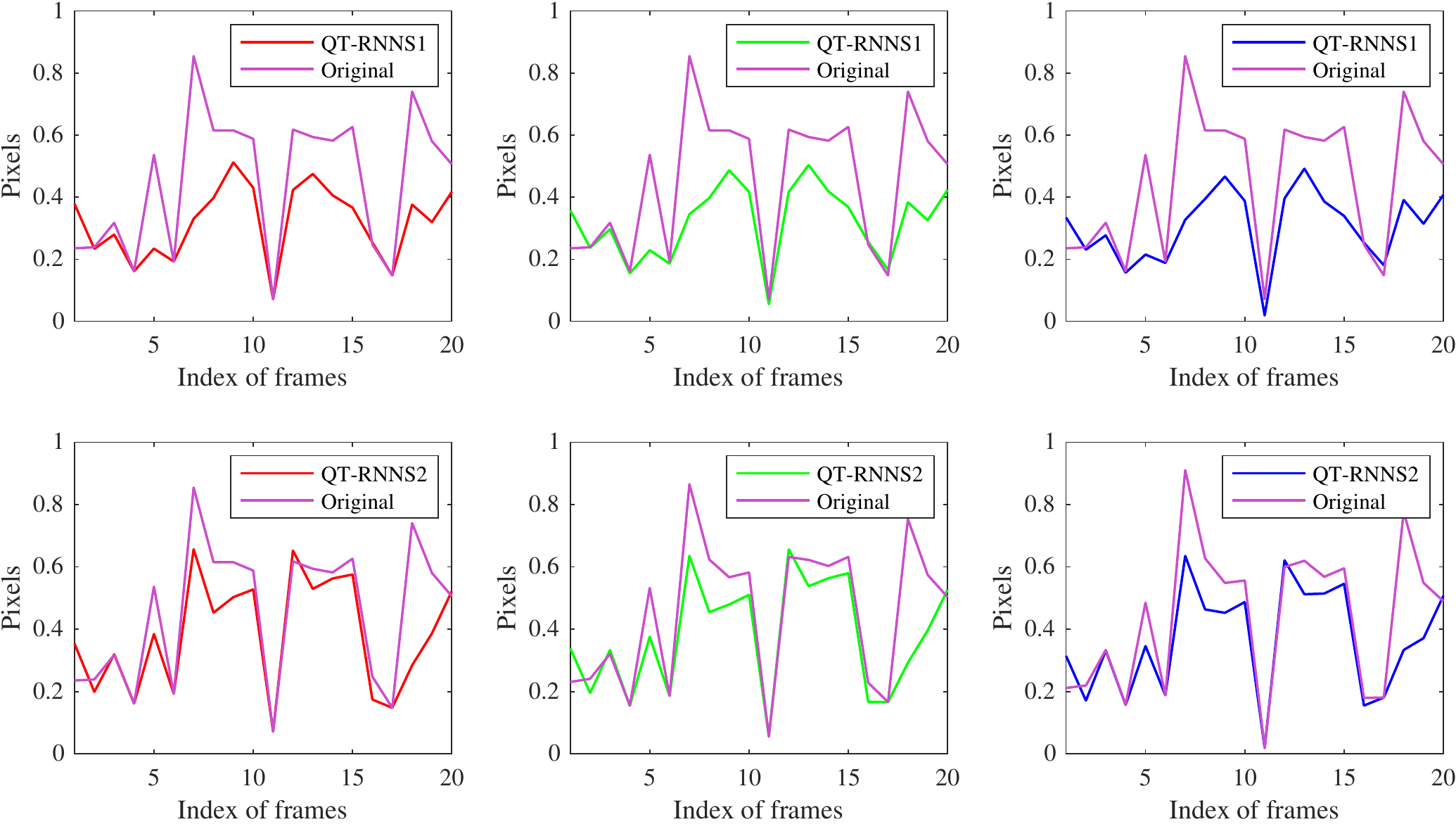}
		\end{minipage}
	}
	\subfigure{
		\begin{minipage}{18cm}
			\includegraphics[height = 4.8cm, width = 8.5cm]{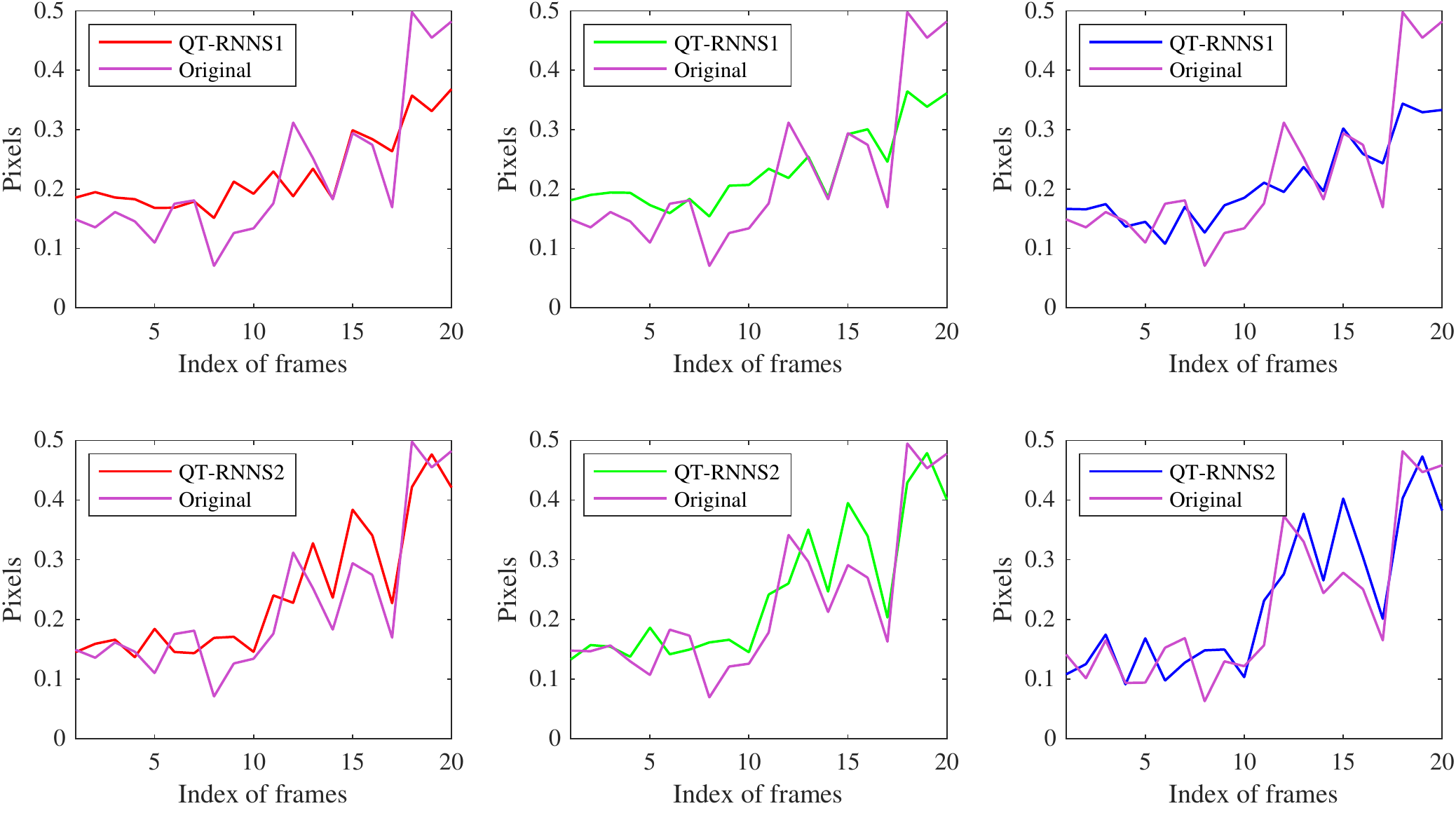}
		\end{minipage}
	}
	\subfigure{
		\begin{minipage}{18cm}
			\includegraphics[height = 4.8cm, width = 8.5cm]{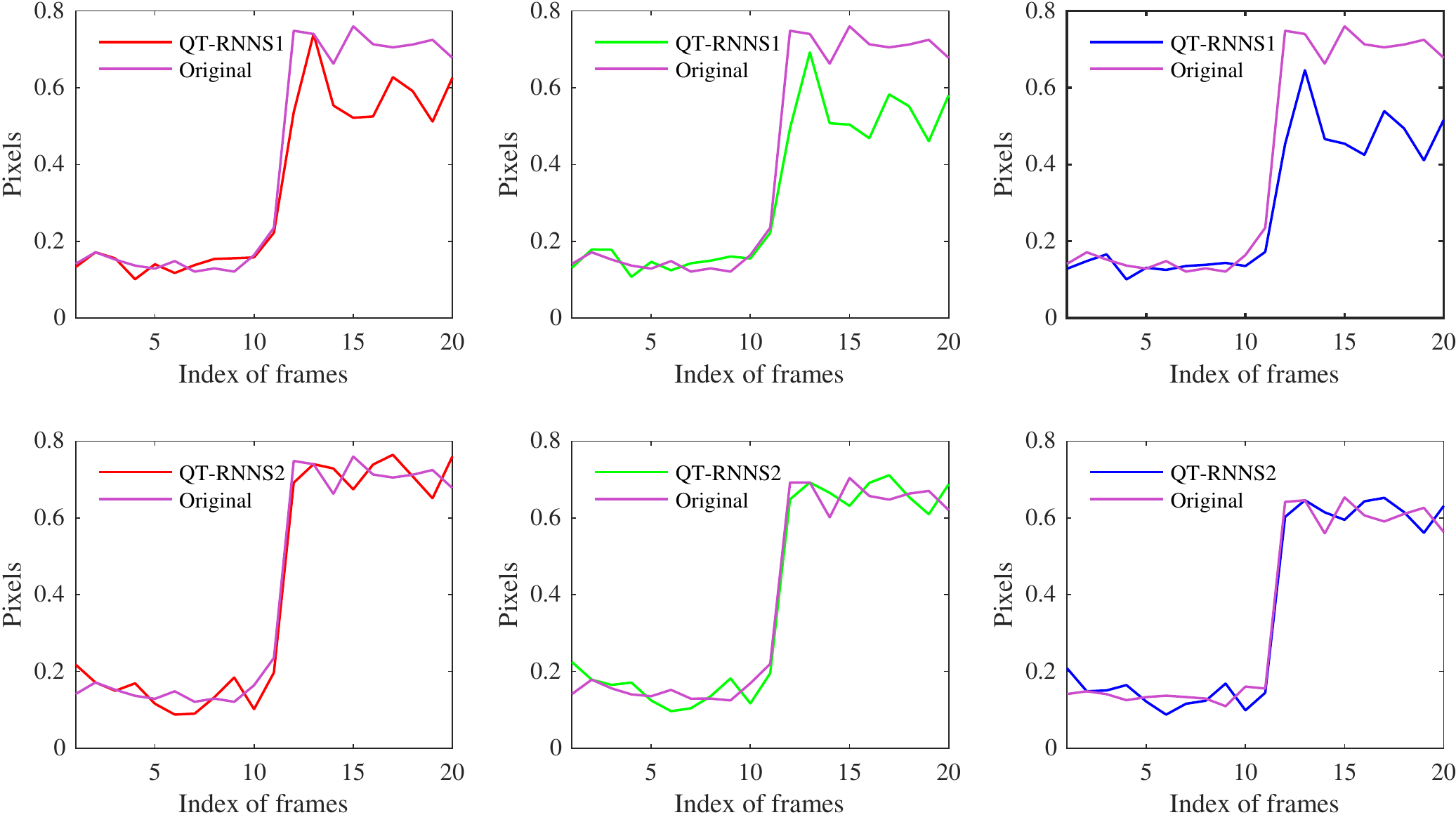}
		\end{minipage}
	}
	\caption{The pixels recovered results of color video \textit{bus} by operating the proposed algorithms with SR$=0.1$. The first two rows are the pixels at the spatial location (206,100), the middle two rows are the pixels at the spatial location (220,300), and the bottom two rows are the pixels at the spatial location (11,16). The first, second, and third columns are the results for the red, green, and blue channels, respectively.}
	\label{b}
\end{figure}

\begin{figure*}[htbp]
	\centering
	\captionsetup{font={footnotesize}}
	\subfigure{
		\begin{minipage}{18cm}
			\centering
			\includegraphics[height = 5cm, width = 18 cm]{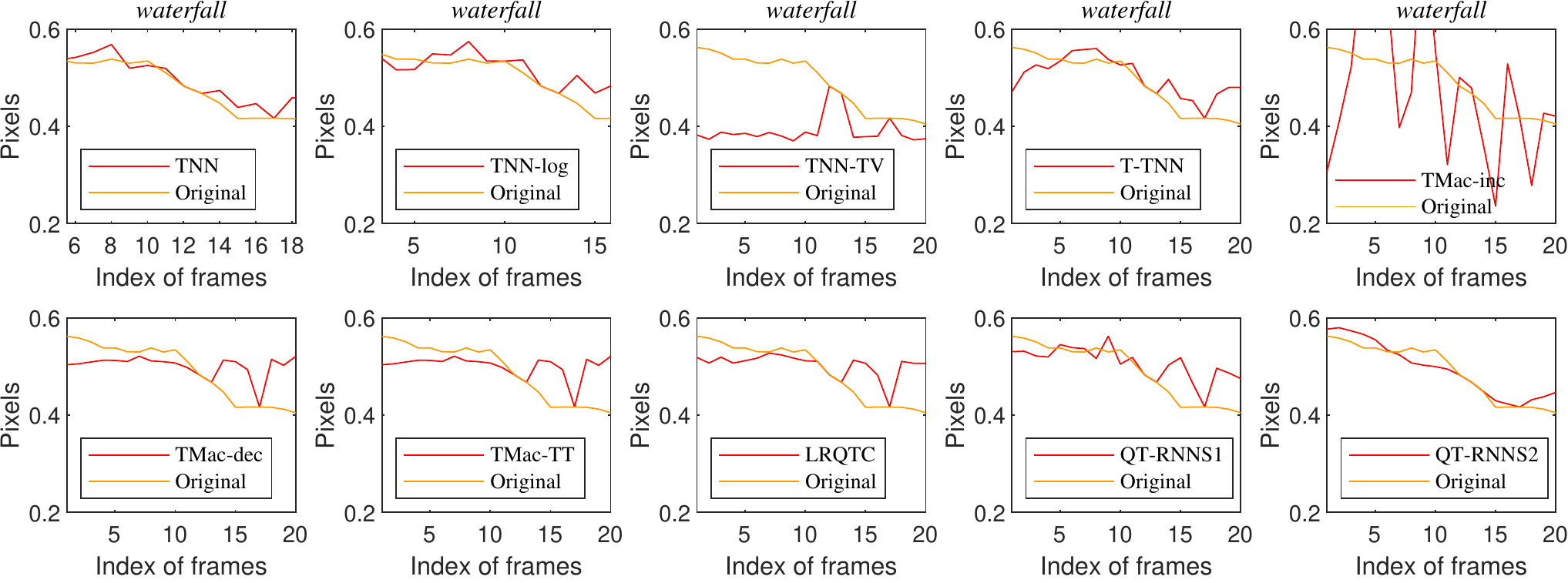}
		\end{minipage}
	}
	\subfigure{
		\begin{minipage}{18cm}
			\centering
			\includegraphics[height = 5cm, width = 18cm]{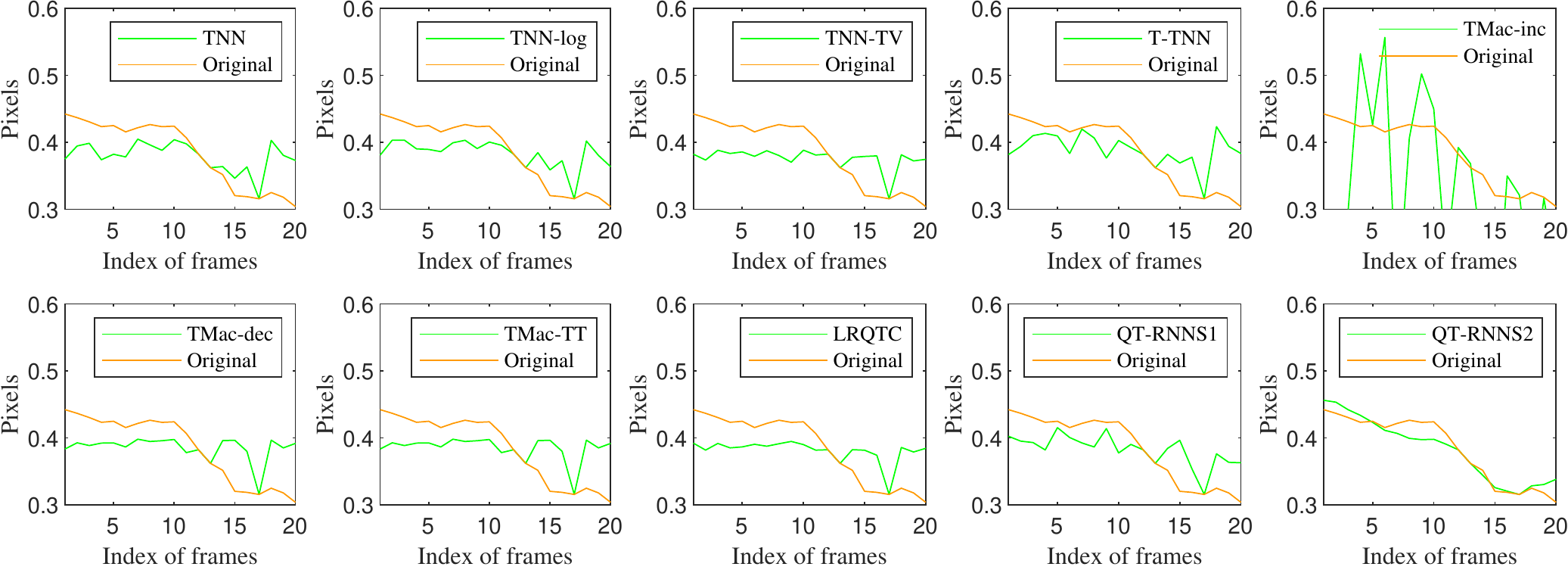}
		\end{minipage}
	}
	\subfigure{
		\begin{minipage}{18cm}
			\centering
			\includegraphics[height = 5cm, width = 18cm]{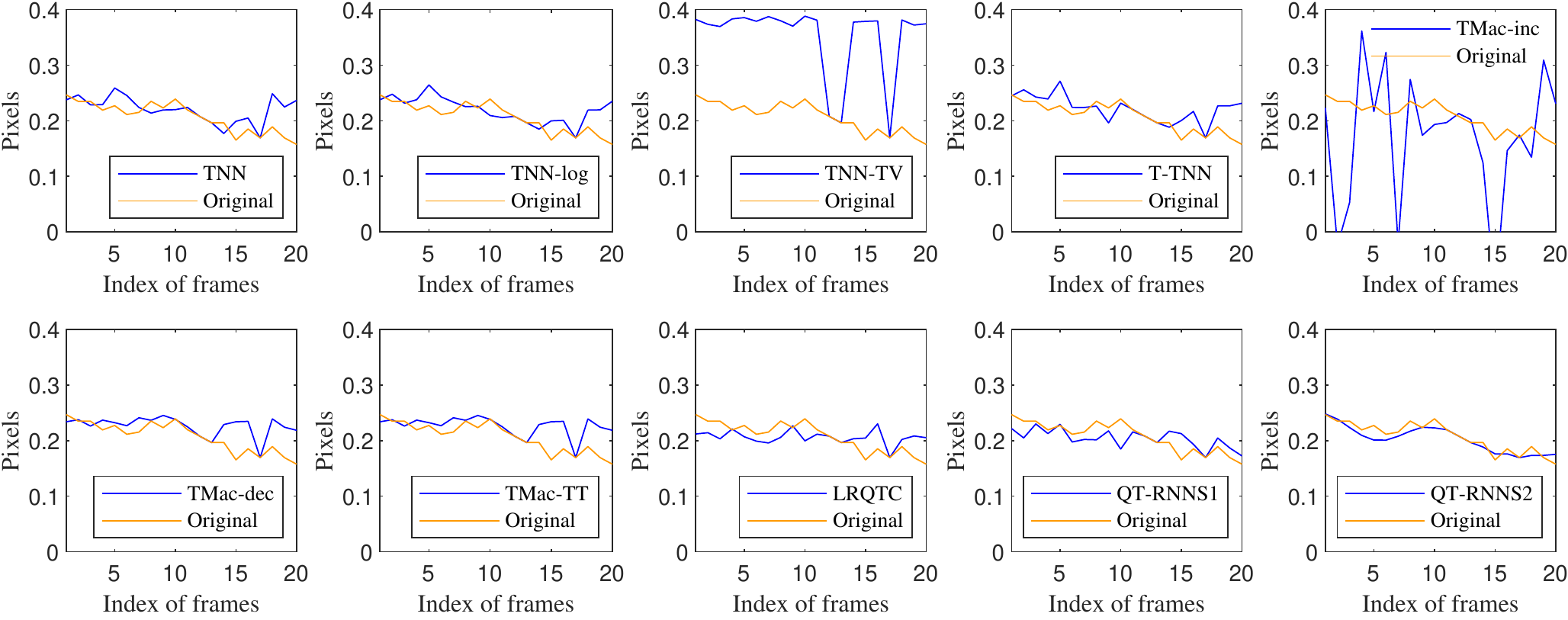}
		\end{minipage}
	}
	\caption{Comparison of the pixels recovered results of different algorithms with SR$=0.1$ at spatial location (153,104) of video \textit{waterfall}. The first two rows are the results of the red channel, the middle two rows are the results of the green channel, and the bottom two rows are the blue channel.}
	\label{w}
\end{figure*}

The visual comparison is shown in Fig.\ref{7b}-Fig.\ref{7s} in case of SR=$0.3$ (taking \textit{bus} and \textit{stefan} as examples). The picture in the green box in the lower right corner is the double magnification of the red box in the picture. As observed in Fig.\ref{7b}, the letters in the picture recovered by QT-RNNS2 are the most distinct. In Fig.\ref{7s},  it is possible to observe the tennis racket Stefan holding by carrying out QT-RNNS1 and QT-RNNS2.

In the case of SR=$0.1$, the visual comparison is shown in Fig.\ref{b1}-Fig.\ref{c}. As can be seen, the results obtained by QT-RNNS2 have generally better visual effects than the recovered results obtained by other compared methods. The proposed method not only retains the global structure of the color video but also recovers its details. For example, it can be clearly observed from Fig. \ref{st},  \textit{stefan}'s shoes can be seen in the recovered result of QT-RNNS2, while the recovery results of other methods are less visible. 

In Fig. \ref{b}, the reconstructed pixel values of the color video \textit{bus} at three spatial locations (206, 100), (220, 300), (11, 16) are shown. The pink curves represent the original pixels, whereas the red, green, and blue curves represent the recovered pixels. 

Further, in Fig. \ref{w}, the reconstructed pixel values of the color video \textit{waterfall} at one spatial location (153, 104) are shown. The yellow curves represent the original pixels, whereas the red, green, and blue curves are the recovered pixels. It can be observed that QT-RNNS2 can produce results that are the closest to the original curve. When the curves of QT-RNNS1 and QT-RNNS2 are compared, it is clear that the logarithmic function captures the rank function more well than other techniques. 

\subsection{Discussions} \label{d}
The following is a summary of the discussion of the above experiment results:
\begin{itemize} 
	\item The developed QT-RNNS1 and QT-RNNS2 methods can achieve such excellent performance because they take into account both the global low rankness of color videos and the sparseness. Moreover, the quaternion tensor-based framework can prevent dimensionality reduction loss.
	\item In comparison to tensor-based methods: the proposed QT-RNNS1 and QT-RNNS2 are based on TQt-SVD which is comparable to tensor SVD in that the structure of color videos is preserved in the whole process. Besides, the simulation results demonstrate that the benefit of utilizing a truncated strategy and a logarithmic function to pursue a more accurate rank approximation is effectively implemented in newly created quaternion-based methods.  
	\item In comparison to quaternion tensor-based methods: QT-RNNS1 and QT-RNNS2 are based on TQt-rank, and further approximation of the rank function improves the accuracy of the recovery results. 
\end{itemize}
\section{Conclusion}
\label{C}
In this paper, a new quaternion tensor recover model was proposed for color video completion. The proposed method is based on TQt-rank, and then utilizes truncation and logarithmic function to obtain a more accurate low-rank restriction. A novel QTDCT was also developed to address the sparseness of color videos in the model. Furthermore, this regularization was induced by ${l_1}$ norm in the proposed model. A two-step optimization with ADMM framework was adopted to optimize the model. Experimental results for color videos recovery demonstrate the superiority of the proposed model. However, there remains significant room for improvement in color video recovery using quaternion tensors, such as introducing other decomposition methods in tensors (ring decomposition, fully-connect tensor network decomposition, etc.) into quaternion tensors, such as more in-depth exploration of the local properties of color video (smoothness, continuity, etc.).

\section*{Acknowledgments}
This work was supported by the University of Macau (MYRG2022-00108-FST), Science and Technology Development Fund, Macao S.A.R (FDCT/0036/2021/AGJ, FDCT0015/2019/ASC).


\bibliographystyle{IEEEtran}
\bibliography{IEEEabrv, mybibfile}

\end{document}